\documentclass[pra,twocolumn,english,groupaddress,floatfix,longbibliography]{revtex4-2}
\usepackage[utf8]{inputenc}

\usepackage{natbib}
\usepackage{graphicx}
\usepackage{xcolor}
\usepackage{bbold}
\usepackage{hhline}
\usepackage{bm}
\usepackage{booktabs}
\usepackage{amsmath}
\usepackage{amssymb}
\usepackage[switch]{lineno}
\usepackage{cuted}

\begin{document}
\title{Wideband Josephson Parametric Isolator}

\author{M. A. Beck} 
\email{matthew.beck@ibm.com}
\author{M. Selvanayagam}
\author{A. Carniol}
\author{S. Cairns}
\author{C. P. Mancini}
\affiliation{IBM Quantum, IBM T.J. Watson Research Center, Yorktown Heights, NY 10598, USA}
\date{\today}

\begin{abstract}

The cryogenic hardware required to build a superconducting qubit based quantum computer demands a variety of microwave components. These elements include microwave couplers, filters, amplifiers, and circulators/isolators. Traditionally
implemented as discrete components, integration of this peripheral hardware, in an effort to reduce overall footprint, thermal load, and added noise, is a key challenge to scaling modern quantum processors with qubit counts climbing over the 100+ mark. Ferrite--based microwave isolators, generally employed in the readout chain to decouple qubits and resonators from  readout electronics, persist as one of the volumetrically largest devices still utilized as discrete components.
Here we present an alternative two--port isolating integrated circuit derived from the DC Superconducting Quantum Interference Device (DC--SQUID).  Non--reciprocal transmission is achieved using the three--wave microwave mixing properties of a flux-modulated DC--SQUID. We show that when multiple DC--SQUIDs are embedded in a multi--pole admittance inverting filter structure, the three--wave mixing derived from the flux pumping of the DC--SQUIDs can provide directional microwave power flow. For a three--pole filter device, we experimentally demonstrate a directionality greater than 15 dB over a 600 MHz bandwidth.

\end{abstract}
\maketitle

\section{INTRODUCTION}

Readout of superconducting quantum processing units (QPU's) is generally performed by dispersively coupling linear microwave resonators to individual qubits \cite{Wallraff:2005aa, Chen:2012aa, Walter2017}. Determination of the qubit states is achieved via the application of a weak (P $\sim$ -120 dBm) microwave probe tone to the linear resonator where the qubit state is mapped to a phase shift of the probe tone. This probe tone and its corresponding phase shift is then amplified, demodulated, and digitized at room temperature.

In many state--of--the--art QPU designs, multiplexed readout of multiple resonators is achieved via a common coupled transmission line \cite{Walter2017,neilSupremacy2018,Lasiq2022}.
In--band noise or signals emanating back towards the QPU from downstream readout electronics can introduce spurious photon populations to the quantum readout resonators leading to excess dephasing \cite{Boissonneault:2009aa, Rosenthal:2021aa}. Modern systems generally decouple the QPU from this downstream noise via commercial ferrite--based cryogenic microwave isolators with single junction isolators generally achieving better than 20+ dB level of in--band isolation. Often,  systems will serially cascade multiple devices in an effort to increase the overall isolation and thus prevent degradation of QPU performance. 

As a consequence of their large physical footprint, placing numerous isolators at the mixing chamber of a dilution refrigerator to support increasing qubit counts is a daunting system integration challenge. In an effort to meet the future demands of larger quantum systems, integrated circuit approaches towards realizing an isolator replacement is an active area of research \cite{Wu:2019aa}. Recently, the inherent non--linear inductance of Josephson junctions (JJs) has been utilized to realize non--reciprocal behaviour \cite{Ranzani:2017aa}. Numerous JJ--based devices have been proposed building off of existing amplifier designs including non--reciprocal devices derived from JPA's \cite{Lecocq:2017aa}, JRM's \cite{Abdo:2019,Abdo:2021,Chien:2020aa,Sliwa2015}, travelling wave devices \cite{Zhang:2021aa,Naghiloo:2021}, and other circuit topologies \cite{Chapman:2017}.  In addition, robust theoretical models of non--reciprocal time--varying Josephson junction based circuits have been developed based on multi--mode modelling and graph theory \cite{Ranzani:2015aa,Naaman:2021, Sliwa2015}.

In this letter, we present our work on the development of cryogenic wideband isolators utilizing SQUIDs as a non--linear mixing element. We present this work in seven sections. Building on the works detailed in \cite{Wu:2019aa, Naaman:2021}, section \ref{Section:CoupledMode} analyzes a generalized case of the parametrically pumped LC circuits presented utilizing coupled--mode theory \cite{Naaman:2021}. We derive analytical expressions for the observed directionality $D$ as a function of the pump strength, pump frequency, and the differential pump phase. In section \ref{section:Circuit}, we demonstrate how a RF flux modulated DC-SQUID gives rise to a spectral impedance matrix where one can directly calculate the three-- and four--wave mixing products. We further show how the employment of 2+ RF flux pumped DC--SQUIDs can give rise to constructive and destructive interference depending on the differential flux modulation phase between two SQUIDs the are capacitively coupled. Section \ref{section:Circuit} concludes by comparing spectral ABCD and harmonic balance simulations of a three--pole admittance inverting filter with embedded flux modulated DC--SQUIDs. Sections \ref{Section:Fab} and \ref{section:Measurements} detail the fabrication and measurements of superconducting wideband parametric isolators utilizing flux modulated DC--SQUIDs, respectively. Sections \ref{Section:Future} and  \ref{Section:Conclusions} describe paths for future work and conclusions. 

\section{Coupled--Mode Isolating Filter}
\label{Section:CoupledMode}
 
\textcolor{black}{Admittance inverting filters are a useful tool in the development of wideband superconducting parametric devices finding use in the development of RF switches \cite{naaman:2016}, amplifiers \cite{Naaman:2019aa}, and IQ mixers \cite{naaman:2017}.}  In this section, we utilize coupled--mode theory \cite{Ranzani:2015aa,Naaman:2021,Alvarez:2019} to investigate the dynamics of a two--pole admittance inverting filter circuit where the resonant frequency \textcolor{black}{of each pole} is modulated via a RF pump. Figure~\ref{Fig:MInverse}(a) displays a diagram of the circuit. We note that the following treatment is agnostic \textcolor{black}{as to the physical implementation of the non-linearity, i.e. whether the shunt inductor or shunt capacitor of the pole \cite{Wu:2019aa} embody the non--linear component subject to modulation. However as described later, we will look at an implementation of a filter where the shunt inductor is replaced with a DC-SQUID. In the following section, we focus on a two--pole admittance inverting filter topology and look to understand the constraints on the pump parameters including the relative phase between pumps, the pump frequency, and the pump amplitude required to realize non--reciprocal transmission.}

\subsection{Determining The Pump Parameters}

\begin{figure}[b] 
    \centering
      \includegraphics[width = \linewidth]{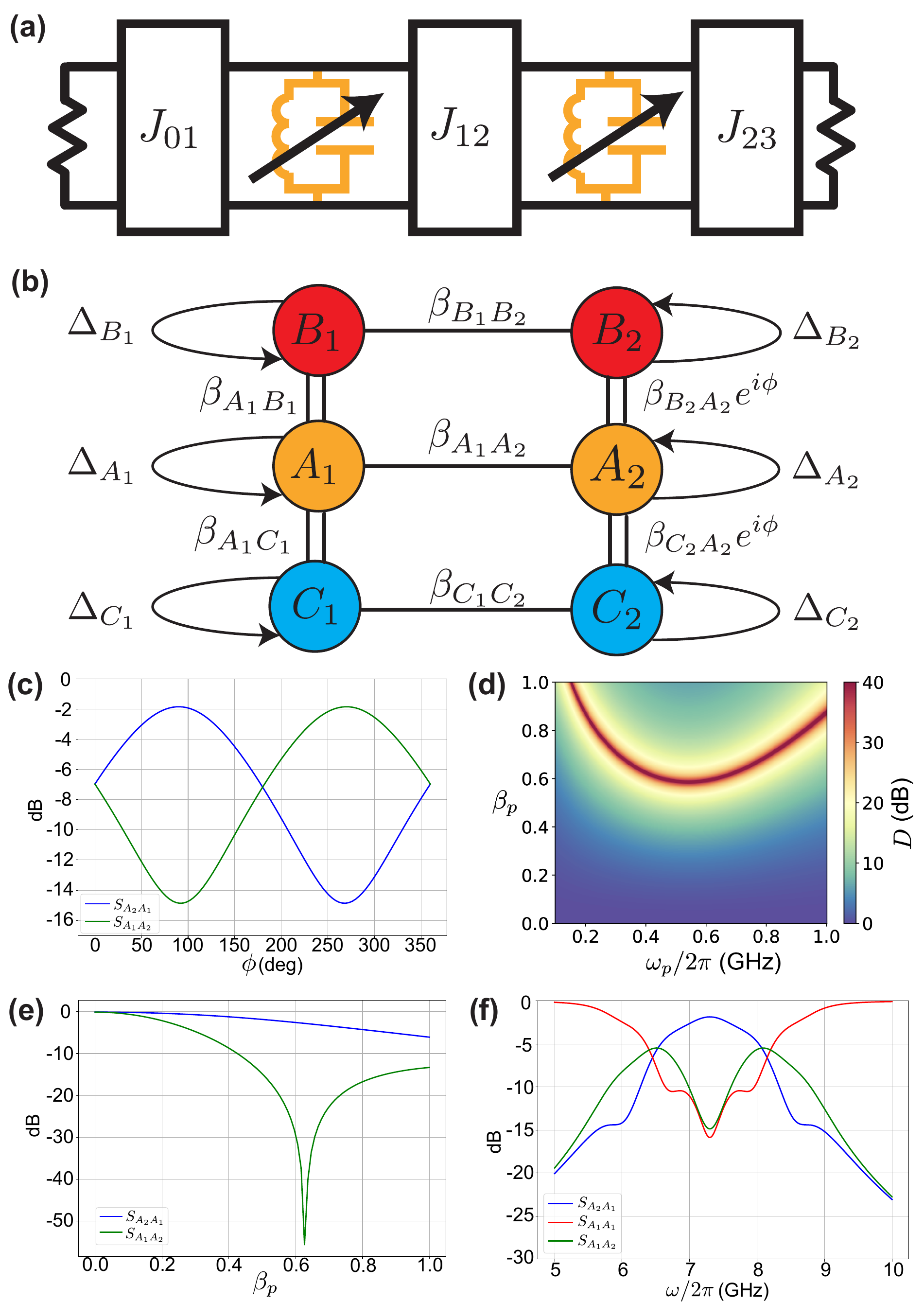}
      \caption{(color online) \textbf{(a)} Two--pole band pass admittance inverter filter diagram with modulated LC resonances.  \textbf{(b)} Coupled mode graph equivalent of \textbf{(a)}. \textbf{(c)} Forward $S_{A_2A_1}$ and reverse $S_{A_1A_2}$ transmission as a function of the differential flux pump phase $\phi$. \textbf{(d)} False color plot of the directionality $D$ as a function of the pump frequency $\omega_p$ and pump strength $\beta_p$ for a phase $\phi = \pi/2$. \textbf{(e)} Forward and reverse scattering parameters as a function of the pump strength $\beta_p$. \textbf{(f)} Forward, reflected, and reverse scattering parameters for the coupled mode diagram in \textbf{(b)} for a pump strength $\beta_p = 0.5$ and differential pump phase $\phi = \pi/2$.}
      \label{Fig:MInverse}
\end{figure}

We employ a two--pole bandpass filter as our model as shown in FIG.~\ref{Fig:MInverse}(a). In order to capture the dynamics arising solely from the first upper and lower sidebands generated due to the pole frequency modulation, an equivalent coupled--mode network comprised of six distinct modes is utilized. A diagram of the network is shown in FIG.~\ref{Fig:MInverse}(b).  The $A$ modes are at the center frequency of the filter $\omega_o$, the $B$ modes are at upper side band $\omega_B = \omega_o + \omega_p$, and the C modes are at the lower side band $\omega_C = \omega_o - \omega_p$, where $\omega_p$ is the pump frequency. Thus the B and C modes are simply the first two sidebands generated via the pole non-linearity which are detuned from the center frequency by the pump. \textcolor{black}{The coupling matrix formulation of the graph in FIG.~\ref{Fig:MInverse} is}
%
\begin{align}\label{coupleMatrix}
\resizebox{0.42\textwidth}{!}{$
    M = \begin{bmatrix}
        \Delta_{A_1} & \beta_{A_1A_2} & \beta_{A_1B_1} & 0 & \beta_{A_1C_1} & 0 \\
        \beta_{A_2A_1} & \Delta_{A_2} & 0 & \beta_{A_2B_2}e^{i\phi} & 0 & \beta_{A_2C_2}e^{-i\phi} \\
        \beta_{B_1A_1} & 0 & \Delta_{B1} & \beta_{B_1B_2} & 0 & 0 \\
        0 & \beta_{B_2A_2}e^{-i\phi} & \beta_{B_2B_1} & \Delta_{B_2} & 0 & 0 \\
        \beta_{C_1A_1} & 0 & 0 & 0 & \Delta_{C_1} & \beta_{C_1C_2} \\
        0 &  \beta_{C_2A_2}e^{i\phi} & 0 & 0 &  \beta_{C_2C_1} & \Delta_{C_2} \\
    \end{bmatrix} \, , $}
\end{align}
\textcolor{black}{where $\Delta_X = 1/\gamma_0\left(\omega_X^s - \omega_X + i\gamma_A/2\right)$
encapsulate the detuning of the mode $\omega_X$ from the applied signal frequency $\omega^s_X$. The $\gamma_x$ term represents the coupling between the mode and the external ports. The leading term $\gamma_0 = \sqrt[N]{\Pi_{n=1}^N \gamma_i}$ is the geometric mean loss rate for the coupled mode system.} The $\beta_{XY}$ terms describe the coupling between modes $X$ and $Y$. Double--line connections represent parametric coupling between modes and single--line connections represent passive coupling \cite{Naaman:2021}. Sidebands linked to the center frequency via parametric couplings are defined by the modulation amplitude and phase. We make the coupling phase explicit via a differential phase offset term $\phi$ between modes $B_2$--$A_2$ and modes $C_2$--$A_2$ so as to provide the required degree of freedom to create constructive and destructive interference. 

To calculate the required differential phase, we take the limit where the input and output coupling are equal $\gamma_i = \gamma_o = \gamma_0$ and the signal frequency is equal to the $A$ mode. In this limit, Eq.~\eqref{coupleMatrix} reduces to
\begin{align}\label{coupleMatrix2}
\resizebox{0.42\textwidth}{!}{$
    M = \begin{bmatrix}
        i/2 & \beta_{c} & \beta_{p} & 0 & \beta_{p} & 0 \\
        \beta_{c} & i/2 & 0 & \beta_{p}e^{i\phi} & 0 & \beta_{p}e^{-i\phi} \\
        \beta_{p} & 0 & i/2 + \omega_p/\gamma_0 & \beta_{c} & 0 & 0 \\
        0 & \beta_{p}e^{-i\phi} & \beta_{c} & i/2 + \omega_p/\gamma_0 & 0 & 0 \\
        \beta_{p} & 0 & 0 & 0 & i/2 - \omega_p/\gamma_0 & \beta_{c} \\
        0 &  \beta_{p}e^{i\phi} & 0 & 0 &  \beta_{c} & i/2-\omega_p/\gamma_0 \\
    \end{bmatrix} \, , $}
\end{align}
where we have set the coupling rates between degenerate frequencies $\beta_{X_1X_2} = \beta_{c}$ and the parametric couplings $\beta_{X_NY_N} = \beta_p$. To calculate the directionality, we invert Eq. \eqref{coupleMatrix2} and take the ratio of the forward and reverse coupling terms.
\begin{subequations}
\begin{align}
    D &= \frac{|S_{A_2A_1}|}{|S_{A_1A_2}|} = \frac{|M_{A_2A_1}^{-1}|}{|M_{A_1A_2}^{-1}|} \\
    &= \frac{|M_a + \zeta \cos(\phi) - \eta \sin(\phi)|}{|M_a + \zeta\cos(\phi) + \eta\sin(\phi)|} \label{Eq:D2} \,
\end{align}
\end{subequations}
where
\begin{align}
    M_a &= -\frac{\beta_{c}}{16} - \frac{\beta_{c}^3}{2} - \beta_{c}^5 \\
    &- \frac{\beta_{c}\omega_p^2}{2\gamma_0^2} + \frac{2\beta_{c}^3 \omega_p^2}{\gamma_0^2}
    - \frac{\beta_{c}\omega_p^4}{\gamma_0^4} \nonumber \\
    \zeta &= \frac{1}{2}\beta_{c}\beta_{p}^2 + 2\beta_{c}^3\beta_p^2
    -2\frac{\beta_{c}\beta_p^2\omega_p^2}{\gamma_0^2} \\
    \eta &= 2\frac{\beta_{c}\beta_p^2\omega_p}{\gamma_0} \, .
\end{align}
Upon inspection of Eq.~\eqref{Eq:D2}, we see that for $\phi = n\pi$ (where $n$ is an integer), $D = 1$ for any and all values related to the pump thus yielding matching forward and reverse scattering amplitudes. When \textcolor{black}{$\phi = (n+1/2)\pi$}, the amount of directionality depends on the ratio $M_a/\eta$ with infinite directionality (complete suppression of $S_{A_1A_2}$) achieved when $M_a/\eta = 1$. \textcolor{black}{For this case, we derive two limits on the pump amplitude as a function of the pump frequency. In both cases we take $\beta_c \ll 1$. In the limit $a\equiv \omega_p/\gamma_0 \ll 1$, the pump amplitude $\beta_p \propto \sqrt{1/a}$ for infinite directionality. In the opposite limit $a \gg 1$, the pump amplitude $\beta_p \propto \sqrt{a^3}$}. 

\textcolor{black}{Given the above, we can also say a few words regarding the insertion loss. In the limit $\beta_c^r\beta_p^s \ll 1$ where ${r+s\geq3}$, we can derive a qualitative relation between the forward transmission and the pump amplitude. In this limit, the forward transmission takes the form 
\begin{align}\label{S21VsPumpAmp}
    |S_{A_2A_1}| \approx \frac{1 + \beta_p^2}{1 + b\beta_p^2} \, ,
\end{align}
where $b$ is a numerical factor greater than 1. With the the pump turned off, $|S_{A_2A_1}| = 1$ yielding unity transmission. In the limit of increasing $\beta_p$, the transmission asymptotically approaches $|S_{A_2A_1}| \rightarrow 1/b$}.

We numerically calculate the forward and reverse scattering parameters via the conversion of the two--pole network in FIG.~\ref{Fig:MInverse}(a) to the equivalent coupled--mode structure in FIG.~\ref{Fig:MInverse}(b). Our model has a center frequency $\omega_A = 7.3$ GHz, bandpass ripple $r = 0.1$ dB and a bandwidth BW = 750 MHz. Figure~\ref{Fig:MInverse}(c) displays $S_{A_1A_2}$ and $S_{A_2A_1}$ as a function of the differential pump phase for a on--resonance signal drive applied to modes $A_1$ and $A_2$ with $\beta_p = 0.5$. At differential phases $\phi = n\pi$ the simulations show $S_{A_1A_2} = S_{A_2A_1}$ resulting in unity directionality as predicted by Eq.~\eqref{Eq:D2}. The maximum difference is seen at $\phi=\pi/2$. Figure~\ref{Fig:MInverse}(d) plots the directionality $D$ in dB in false color as a function of the pump frequency $\omega_p$ and pump strength $\beta_p$. The color bar is artificially cutoff at 40 dB. The non--linear red line displays when the denominator of Eq.~\eqref{Eq:D2} goes to 0 resulting in full suppression of $S_{A_1A_2}$. Figure~\ref{Fig:MInverse}(e) plots $S_{A_2A_1}$ and $S_{A_1A_2}$ as a function of the pump strength $\beta_p$ for a pump frequency $\omega_p/2\pi = 700$ MHz. The simulations show that while exponential suppression of $S_{A_1A_2}$ is possible as one approaches a pump amplitude $\beta_p \approx 0.62$, a trade--off is made between the maximum suppression of $S_{A_1A_2}$ and the \textcolor{black}{additive} insertion loss in $S_{A_2A_1}$ \textcolor{black}{stemming from the stiffened pump shown shown in Eq.~\eqref{S21VsPumpAmp}}. This additional insertion loss is due to imperfect constructive interference resulting in power lost to the sideband mode external couplings. Figure~\ref{Fig:MInverse}(f) plots the forward, reflected, and reverse scattering parameters for the coupled--mode filter for a pump amplitude $\beta_p = 0.5$. We see a suppression of $S_{A_1A_2}$ of approximately 15 dB with a corresponding insertion loss of 2 dB. An important aspect of this plot is that it predicts that the system stays matched to better than 10 dB indicating that the nominal matched response of the underlying linear filter is maintained while under modulation. This allows us to treat the pump as a perturbative parameter to the overall circuit performance.

\subsection{Bounding The Pump Frequency}

Referring again to FIGS.~\ref{Fig:MInverse}(a--b) and the resulting calculation, when one inserts the forms for the lower and upper sideband frequencies for perfect tuning, all terms on the diagonal of of Eq.~\eqref{coupleMatrix} become functions of the applied signal frequency $\omega^s_{A}$ and the $A$ mode frequency $\omega_A$ \textit{only}, providing little insight into exactly what pump frequency one should apply in an integrated broadband filter circuit. For discrete modes, the answer is simply that the pump should be the difference between the signal and respective sideband mode frequencies. In a continuous band circuit, what exactly constitutes a discrete mode is slightly more complicated.

To gain insight into what sets the pump frequency in a broadband circuit, we take the case where the applied signal frequency is at the center of the filter band, $\omega^s_A=\omega_o$.  We also define the isolation bandwidth $\delta\omega$ centered on $\omega_0$ which represents the frequency bandwidth of the desired directionality for the circuit under modulation. Note that this is separate from the linear filter bandwidth BW and in general $\delta\omega \leq$ BW.  Two points in the continuous signal frequency band are considered, $\omega_o+\delta\omega/2$ and $\omega_o-\delta\omega/2$ which are the upper and lower edges of the isolation bandwidth. Under modulation both of these frequencies generate lower and upper sidebands, $\omega_{B+},\omega_{C+}, \omega_{B-}, \omega_{C-}$ for a given pump frequency $\omega_P$. These are defined as
\begin{align}
    \omega{_B+} &= \omega_A + \frac{\delta\omega}{2} + \omega_P,  \label{omegaB+} \\
    \omega{_C+} &= \omega_A + \frac{\delta\omega}{2} - \omega_P,  \label{omegaC+}\\
    \omega{_B-} &= \omega_A - \frac{\delta\omega}{2} + \omega_P,  \label{omegaB-} \\
    \omega{_C-} &= \omega_A - \frac{\delta\omega}{2} - \omega_P.  \label{omegaC-}
\end{align}
Depending on the value of $\omega_p$, only $\omega_{B-}$ and $\omega_{C+}$ will fall into the $\omega_A\pm\delta\omega/2$ band while $\omega_{B+}$ and $\omega_{C-}$ always exist outside of it.  By bounding $\omega_P$ such that $\omega_{B+}$ and $\omega_{C-}$ land outside of $\omega_A\pm\delta\omega/2$ we can get an approximate value for the pump frequency. Subtracting Eq. \eqref{omegaC+} from Eq.~\eqref{omegaB-} yields
\begin{align}\label{BWRelation}
   \omega_{B-} - \omega_{C+} = 2\omega_P -\delta\omega \geq \delta\omega, \\
   \omega_P \geq \delta\omega
\end{align}
By setting the pump frequency larger than the desired isolation bandwidth $\delta\omega$, we ensure that no side band frequencies arising from the modulation lie within the isolation band.  Further, we can approximately bound the pump frequency from above by noting that for frequencies $\omega_{B+}$ and $\omega_{C-}$, which would lie outside the filter passband for large $\omega_P$, their participation via their coupling to the filter is greatly reduced once largely detuned from the modes of the filter \cite{Naaman:2021}. Thus, in general, keeping $\omega_P\leq$ BW suffices as a good rule, though we note that this is not a hard bound. 

In practice, once a desired isolation bandwidth is chosen, flexibility exists in choosing the pump frequency.
A good rule-of-thumb is to have $\delta\omega \lessapprox \omega_P \lessapprox$ BW with the goal to maximize $\delta\omega$.  The smaller that $\delta\omega$ can be set with respect to the underlying linear filter bandwidth, the larger freedom there is in choosing $\omega_P$.

\section{Superconducting Circuit Implementation}\label{section:Circuit}

With the above coupled--mode treatment detailing the rise of non--reciprocity in a modulated two--pole LC network, we propose the use of DC--SQUIDs as the non--linear mixing element. In this section, we describe how, under the application of RF flux, DC-SQUIDs give rise to three or four--wave mixing and detail how the application of 2+ RF flux pumped DC--SQUIDs can give rise to directionality in a two port circuit. We conclude the section comparing the results of a spectral ABCD matrix calculation with that of a numerical harmonic balance simulation for a three--pole admittance inverting filter with DC--SQUIDs embedded in each of the shunt LC poles.

\subsection{DC--SQUID Mixer Theory}
A DC--SQUID is comprised of the parallel combination of two Josephson junctions in a superconducting loop \cite{Cantor:2004aa}. The DC--SQUID possesses an inherent inductance with a non--linear dependence on the applied flux through the loop. The form of the inductance is:
\begin{equation}\label{eq:1}
    L_\text{SQ} = \frac{\Phi_0}{2\pi I_c \cos(\pi \Phi_A / \Phi_0)},
\end{equation}
where $\Phi_0 \equiv h/2e$ is the superconducting magnetic flux quantum, $I_c \equiv 2\times I_{c0}$ is twice the critical current ($I_{c0}$) of the individual JJs, and $\Phi_A$ is the applied flux through the DC--SQUID loop. Setting $\pi \Phi_A/\Phi_0 = \beta + \alpha \cos(\omega_m t + \phi)$ where $\beta$ and $\alpha$ are the DC and RF flux amplitudes, $\omega_m$ is the modulation frequency and $\phi$ is the modulation phase, we Taylor expand the $\cos(\pi \Phi_A/\Phi_0)$ about $\beta$ yielding
\begin{align} \label{Eq:SQUIDExp}
    \cos(\pi \Phi_A/\Phi_0) \approx \cos(\beta) &- \sin(\beta)\left(\frac{\pi \Phi_A}{\Phi_0}- \beta\right) \\
    &- \frac{\cos(\beta)}{2}\left(\frac{\pi \Phi_A}{\Phi_0} - \beta\right)^2 \, . \nonumber
\end{align}
Making the substitution $\pi\Phi_A/\Phi_0 - \beta = \alpha\cos(\omega_m t + \phi)$, Eq. \eqref{Eq:SQUIDExp} can be rewritten as
\begin{align}
    \cos(\pi \Phi_A/\Phi_0) \approx \cos(\beta) & \left[ 1 - \tan(\beta)\alpha \cos(\omega_m t + \phi) \right. \\ 
    & \left.- \alpha^2\cos^2(\omega_m t + \phi)/2\right] \, . \nonumber
\end{align}
In the small RF pump limit $\alpha \rightarrow 0$, Eq. \eqref{eq:1} becomes
\begin{align}
    L_\text{SQ}(t) \approx \frac{\Phi_0}{4\pi I_{c0} \cos(\beta)} [1 &+ \tan(\beta)\alpha \cos(\omega_m t + \phi) \\ 
    &+ \alpha^2 \cos^2(\omega_m t + \phi)/2] \nonumber
\end{align}
Finally, expanding the $\cos$ and $\cos^2$ terms into their respective exponential forms yields
\begin{align}\label{Eq:FullExpand}
     L_\text{SQ}(t) \approx &\frac{\Phi_0}{4\pi I_{c0} \cos(\beta)} \times \\ &\left[1 + \frac{\alpha^2}{4}
     + \frac{\tan(\beta)\alpha}{2} \left(e^{i\omega_m t +\phi} + e^{-i\omega_mt - \phi}\right) \right. \nonumber \\
    &\left. +  \frac{\alpha^2}{8}\left(e^{2i\omega_m t + 2\phi} + e^{-2i\omega_m t - 2\phi} \right) \right] \, . \nonumber
\end{align}
The first two terms in Eq.~\eqref{Eq:FullExpand} are the DC modulated and RF modulated terms of the SQUID inductance, respectively. The third and fourth terms are the three--wave mixing terms which modulate  $\omega \rightarrow \omega \pm \omega_m$. The fourth and fifth terms, whose amplitude depend solely on the normalized RF drive strength $\alpha$, give rise to four--wave mixing which take $\omega \rightarrow \omega \pm 2\omega_m$. These modulation tones can drive either amplification or frequency conversion processes depending on their relation to the signal frequency. For the application in question, we will focus mainly on frequency conversion processes but note that this derivation can be generalized to either.

In the limit that the signal frequency $\omega' \gg \omega_m$, one can write the voltage/current relationship of a DC--SQUID as
\begin{align}\label{eq:4}
    V(t) &= L_\text{SQ}(t)\frac{d I(t)}{dt} \, ,
\end{align}
where we have taken the current to be of the form ${I(t) = I_0\exp{[i\omega' t]}}$. Taking the Fourier Transform $\mathcal{F}$ of both sides results in
\begin{equation}\label{eq:5}
    \mathcal{F}\{V(t)\} = i\omega' \mathcal{F}\{L_\text{SQ}(t)\}\ast\mathcal{F}\left \{I(t)\right \} \, .
\end{equation}
This results in a frequency domain voltage/current relationship given by
\begin{subequations}
    \begin{align}\label{eq:6a}
        V(\omega)&=   L_\text{SQ}^0 \left[ \gamma\delta(\omega' - \omega) \right. \\
        &+ \left. \eta_{+} \delta(\omega' - \omega + 2\omega_m) \right. \nonumber \\
        &+ \left. \eta_{-} \delta(\omega'-\omega - 2\omega_m) \right. \nonumber \\
        &+ \left. \kappa_{+} \delta(w'- \omega + \omega_m) \right. \nonumber \\
        &+ \left. \kappa_{-}\delta(\omega' - \omega - \omega_m)\right] \ast i \omega' I(\omega') , \nonumber 
    \end{align}
    \begin{align}\label{eq:6b}
        V(\omega)&=   i\omega L_\text{SQ}^0 \gamma I(\omega) \\
        &+  i\left[\omega+2\omega_m\right] L_\text{SQ}^0  \eta_{+} I(\omega+2\omega_m) \nonumber \\
        &+  i\left[\omega-2\omega_m\right] L_\text{SQ}^0  \eta_{-}I(\omega-2\omega_m)  \nonumber \\ 
        &+ i\left[\omega+\omega_m\right] L_\text{SQ}^0 \kappa_{+} I(\omega+\omega_m)   \nonumber \\
        &+ i\left[\omega-\omega_m\right] L_\text{SQ}^0 \kappa_{-} I(\omega-\omega_m) , \nonumber
    \end{align}
\end{subequations}
where ${L_\text{SQ}^0 = \Phi_0 / 4\pi I_{c0} \cos(\beta)}$, ${\gamma = 1 + \alpha^2/4}$, ${\eta_{\pm} = \alpha^2 e^{\pm i2\phi}/4}$, and ${\kappa_\pm = \tan(\beta)\alpha e^{\pm i\phi}/2}$. The Dirac delta functions in Eq. \eqref{eq:6a} form the connection between the signal and the sidebands produced from the mixing terms. From this, we can formulate the DC--SQUID spectral impedance matrix $\mathbf{Z_\text{SQ}}$ where the voltage, current, and impedance are defined to account for $\pm 2\omega_m$ and $\pm \omega_m$ modulation products resulting from the Taylor expansion of Eq. \eqref{eq:1}.
\begin{equation}
 \begin{bmatrix} 
    V(\omega-2\omega_m) \\ V(\omega-\omega_m) \\ V(\omega) \\ V(\omega+\omega_m) \\ V(\omega+2\omega_m)  \end{bmatrix} = \mathbf{Z_{\text{SQ}}}(\omega) \begin{bmatrix} I(\omega-2\omega_m) \\ I(\omega-\omega_m) \\ I(\omega) \\ I(\omega+\omega_m) \\ I(\omega+2\omega_m)  \end{bmatrix}
\end{equation}
%

Equation \eqref{eq:squidZ} gives the spectral impedance matrix $\mathbf{Z_{\text{SQ}}}(\omega)$ for a single DC--SQUID \cite{Sundqvist:2014} where ${\omega_{n\pm} = \omega \pm n\omega_m}$. Crucially, Eq. \eqref{eq:squidZ} demonstrates that the conversion amplitudes and associated phases between modulation products are \textit{asymmetric} with respect to the conversion from one sideband to another.

\begin{align}\label{eq:squidZ}
         &\mathbf{Z_{\text{SQ}}}(\omega) =  i L_\text{SQ}^0 \times \nonumber \\
            &\begin{bmatrix}
		        \gamma\omega_{2-} & \kappa_-\omega_{2-} & \eta_-\omega_{2-} & 0 & 0\\
		        \kappa_+\omega_{1-} & \gamma\omega_{1-} & \kappa_-\omega_{1-} & \eta_-\omega_{1-} & 0\\
		        \eta_+ \omega & \kappa_+ \omega & \gamma \omega  & \kappa_-\omega & \eta_- \omega \\
		        0 & \eta_+\omega_{1+} & \kappa_+\omega_{1+}  & \gamma \omega_{1+} & \kappa_-\omega_{1+} \\
		        0 & 0 & \eta_+\omega_{2+} & \kappa_+\omega_{2+}  & \gamma \omega_{2+}
	        \end{bmatrix}
\end{align}

\begin{figure}[t] 
    \centering
      \includegraphics[width = \linewidth]{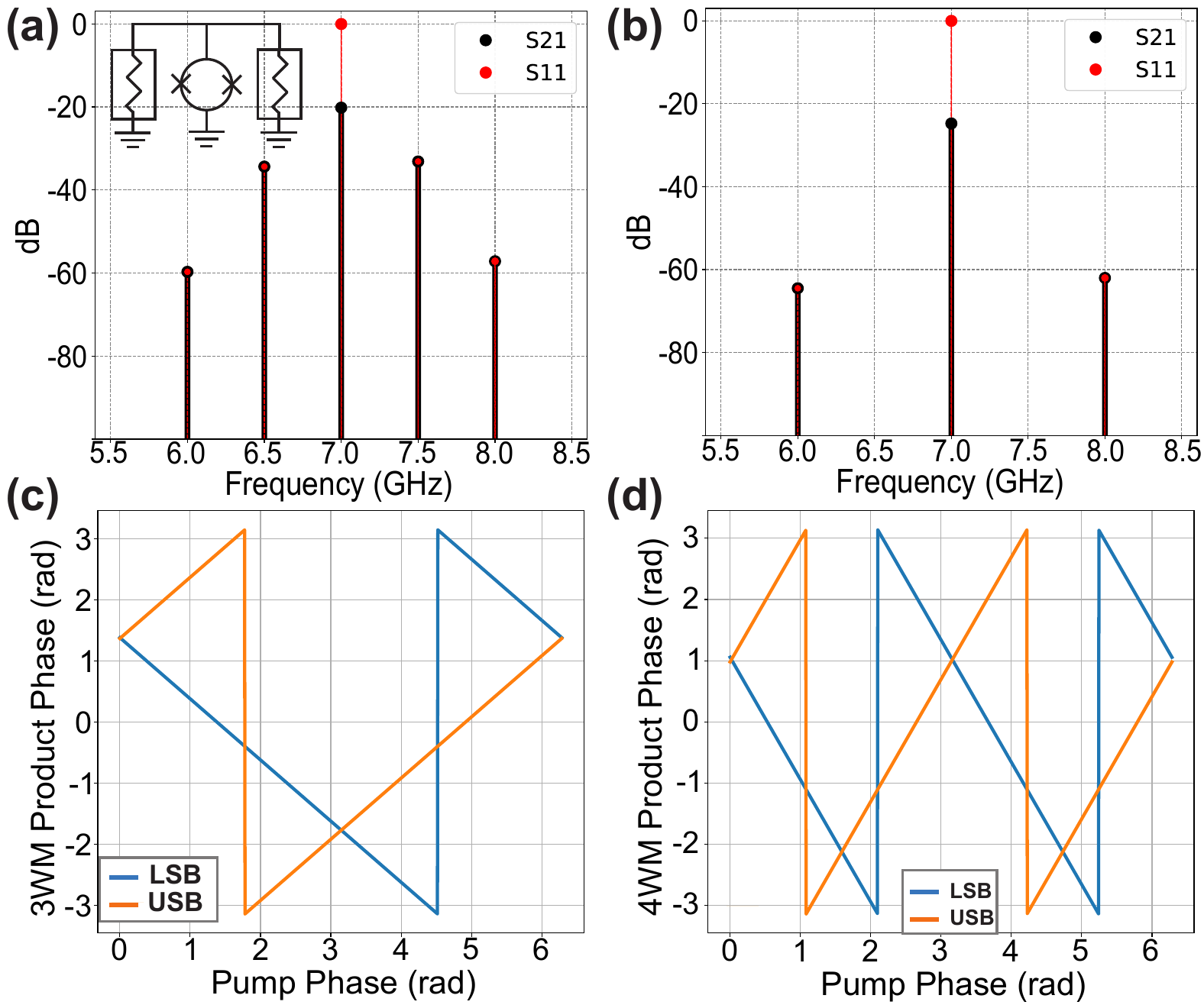}
      \caption{(color online) \textbf{(a)} The forward and reflected scattering spectrum for a DC+RF flux biased DC-SQUID with $\alpha = 0.1\pi$ and $\beta = 0.3\pi$ and a modulation frequency $\omega_m/2\pi = 500$ MHz. (inset) Circuit diagram for the two--port DC--SQUID calculation. \textbf{(b)} Forward and reflected scattering spectrum for a RF flux biased DC SQUID with $\alpha = 0.1\pi$ and a modulation frequency of $\omega_m/2\pi = 500$ MHz. \textbf{(c)} The transmitted phase for the first three--wave mixing (3WM) upper and lower sideband as a function of RF pump phase. \textbf{(d)} The transmitted phase for the first four--wave mixing (4WM) upper and lower sideband as a function of RF pump phase.}
      \label{Fig:3_4WMP}
\end{figure}

\subsection{Two--Port S--parameters of a DC--SQUID}

With the spectral impedance matrix for a DC--SQUID now defined, the scattering parameters of a shunted DC--SQUID can be calculated. The basic circuit is illustrated in the inset of FIG. \ref{Fig:3_4WMP}(a) where a DC--SQUID under RF modulation is shunted to ground between two physical 50--ohm ports. These ports are coupled to the SQUID at all frequencies. The spectral $\mathbf{ABCD}$ matrix for a shunt DC--SQUID is defined to be
\begin{equation}
    \overline{\mathbf{A}}_{\text{SQ}} \equiv
    \begin{bmatrix}
        \mathbf{A} & \mathbf{B} \\
        \mathbf{C} & \mathbf{D} \\
    \end{bmatrix} = 
    \begin{bmatrix}
        \mathbf{I}_{n} & \mathbf{0}_{n} \\
        \mathbf{Z^{-1}_{\text{SQ}}}(\omega) & \mathbf{I}_{n}
    \end{bmatrix} \, ,
\end{equation}
where $\mathbf{I}_{n}$ is the $n\times n$ identity matrix and $\mathbf{0}_{n}$ is a $n \times n$ matrix comprised of all zeroes. The resulting S--parameters are then calculated as \cite{Faria1993,Wu:2019aa}
\begin{subequations}
    \begin{align}\label{eq:SpectralSParams}
        \mathbf{S}_{11} &= \mathbf{I}_{n} - 2\left[\mathbf{I}_{n} + (\mathbf{A}Z_0 + \mathbf{B})\right. \\ &\left.\hspace{10mm}\times (\mathbf{C}Z_0^2 + \mathbf{D}Z_0)^{-1}\right]^{-1} \nonumber \\
        \mathbf{S}_{21} &= 2\left[\mathbf{A} + \mathbf{B}/Z_0 + \mathbf{C}Z_0 + \mathbf{D}\right] \\
        \mathbf{S}_{12} &= 2[\mathbf{I}_{n} + (\mathbf{A}Z_0 + \mathbf{B})(\mathbf{C}Z_0 + \mathbf{D})^{-1}/Z_0 \\
        &\times (\mathbf{A} - (\mathbf{A}Z_0 + \mathbf{B})(\mathbf{C}Z_0 + \mathbf{D})^{-1}\mathbf{C}] \nonumber \\
        \mathbf{S}_{22} &= \mathbf{I}_{n} - 2\left[\mathbf{I}_{n} + (\mathbf{A}Z_0 + \mathbf{C}Z_0^2)^{-1}\right. \\ &\left.\hspace{10mm}\times (\mathbf{B} + \mathbf{D}Z_0)\right]^{-1} \nonumber \, ,
    \end{align}
\end{subequations}
where each entry in the respective spectral S--parameter matrix is defined as
\begin{align}
    S_{ij}^{np} = \frac{b_i(\omega +n\omega_m)}{a_j(\omega+p\omega_m)}\, .
\end{align}
The terms $b_i$ and $a_j$ are the voltage waves entering and leaving ports $i$ and $j$ at their respective frequencies defined by the indices $n$ and $p$. Note that these S--parameter definitions are implicitly functions of the DC bias and RF modulation amplitude applied to the DC--SQUID and require recalculation at every bias point. We note that this definition is similar to what is discussed in \cite{Kaidong:2022aa} in the context of parametric circuits and elsewhere in the literature for non--linear microwave devices in general.

We calculate the S--parameters for a DC--SQUID modulated at the pump tone frequency ${\omega_m/2\pi = f_m = 500}$ MHz with an RF signal tone ${f_s = 7}$ GHz. The critical current of the SQUID JJs is $I_{c0}= 5\, \mu$A and the RF amplitude $\alpha=0.1\pi$. Figures \ref{Fig:3_4WMP}(a) and \ref{Fig:3_4WMP}(b) display the calculated S--parameters in the three--wave ($\beta = 0.3\pi$) and four--wave ($\beta = 0$) mixing cases, respectively.

Having an imaginary admittance, the DC--SQUID interrupts the impedance match between the ports resulting in reduced transmission. However, the three-- and four--wave mixing processes are clearly evident in the spectrum of transmitted and reflected S-parameters. Figures \ref{Fig:3_4WMP}(c) and \ref{Fig:3_4WMP}(d) display the simulated first upper and lower sideband transmitted phase as a function of the modulation pump tone phase for the three-- and four--wave mixing cases, respectively. A well--defined dependency of the transmitted sideband tone phase on the pump tone phase can be clearly seen in both the three-- and four--wave mixing cases.
It is this phase dependency that we aim to utilize in a multi--SQUID architecture to achieve non--reciprocity.

\begin{figure}[t] 
    \centering
      \includegraphics[width = \linewidth]{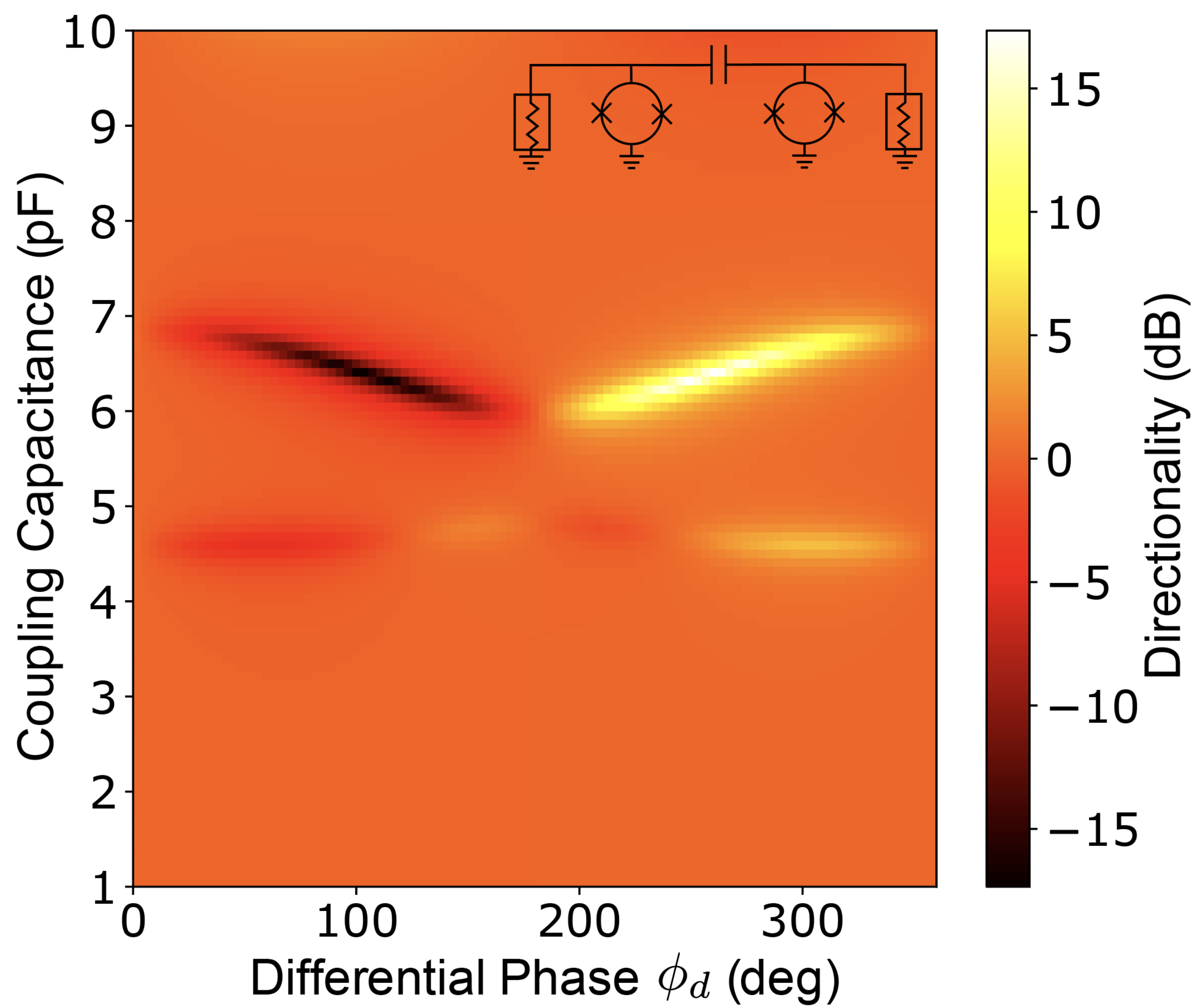}
      \caption{(color online) The directionality of a circuit consisting of two DC-SQUIDs coupled by a capacitor, $C$, embedded between two 50 Ohm terminals. See (inset) for circuit schematic. Each DC-SQUID is modulated with the same parameters in FIG.~\ref{Fig:3_4WMP} with the exception of a differential phase, $\phi_d = \phi_2 - \phi_1$ between each DC-SQUID. The plot defines a false color plot of the directionality $D$ verse the differential phase and coupling capacitance. For specific values of $C$ and $\phi_d$, the directionality is non--zero. This illustrates that to achieve non-reciprocity we need multiple (2+) DC-SQUIDS with a relative pump phase and an appropriate impedance coupling the DC-SQUIDS.
      }
      \label{Fig:Coupled_SQUID}
\end{figure}

\subsection{Generating directionality via use of multiple SQUIDs}

We further extend this model to the two DC--SQUID circuit shown in the inset of FIG. \ref{Fig:Coupled_SQUID}. The circuit consists of two DC--SQUIDS coupled by a series capacitor embedded on either side by 50--Ohm ports. We calculate the measure of non-reciprocity of this circuit as the directionality D defined as
\begin{equation}
    \text{D} =|S^{00}_{21}|/|S^{00}_{12}| \, .
\end{equation}
Each DC-SQUID is driven at the same RF signal frequency $\omega_s$ and modulated with the same parameters shown in FIG.~\ref{Fig:3_4WMP} with the exception of a differential phase, $\phi_d \equiv \phi_2 - \phi_1$, where $\phi_j$ is the phase of the RF tone applied to SQUID $j$. The coupling capacitor is also varied as a way of sweeping the coupling impedance between the two DC-SQUIDS.  FIG.~\ref{Fig:Coupled_SQUID} displays the directionality D as a function of $\phi_d$ and coupling capacitance. It can be seen that the directionality is non--zero for particular sets of coupling capacitance values and phases. 

When cascaded, The DC--SQUIDs serve to mix the signal frequency to the upper and lower sidebands (as shown in FIG.~\ref{Fig:3_4WMP}) and back but with a phase offset all the while the coupling capacitance serves to couple and phase delay the remaining unmodulated signal frequency between ports. In tandem, these two processes can produce constructive or destructive interference resulting in forward (D$>1$) or reverse (D$<1$) directionality. \textcolor{black}{While this simple circuit shows that directionality can be achieved with as little as three circuit elements, without appropriate engineering of their values, a broadband match between the ports cannot be maintained. With this is mind, we chose to embed the SQUIDs into an admittance inverting filter topology to allow for design flexibility.}

\subsection{ABCD--Harmonic Balance Results}

\begin{figure}
    \centering
      \includegraphics[width = \linewidth]{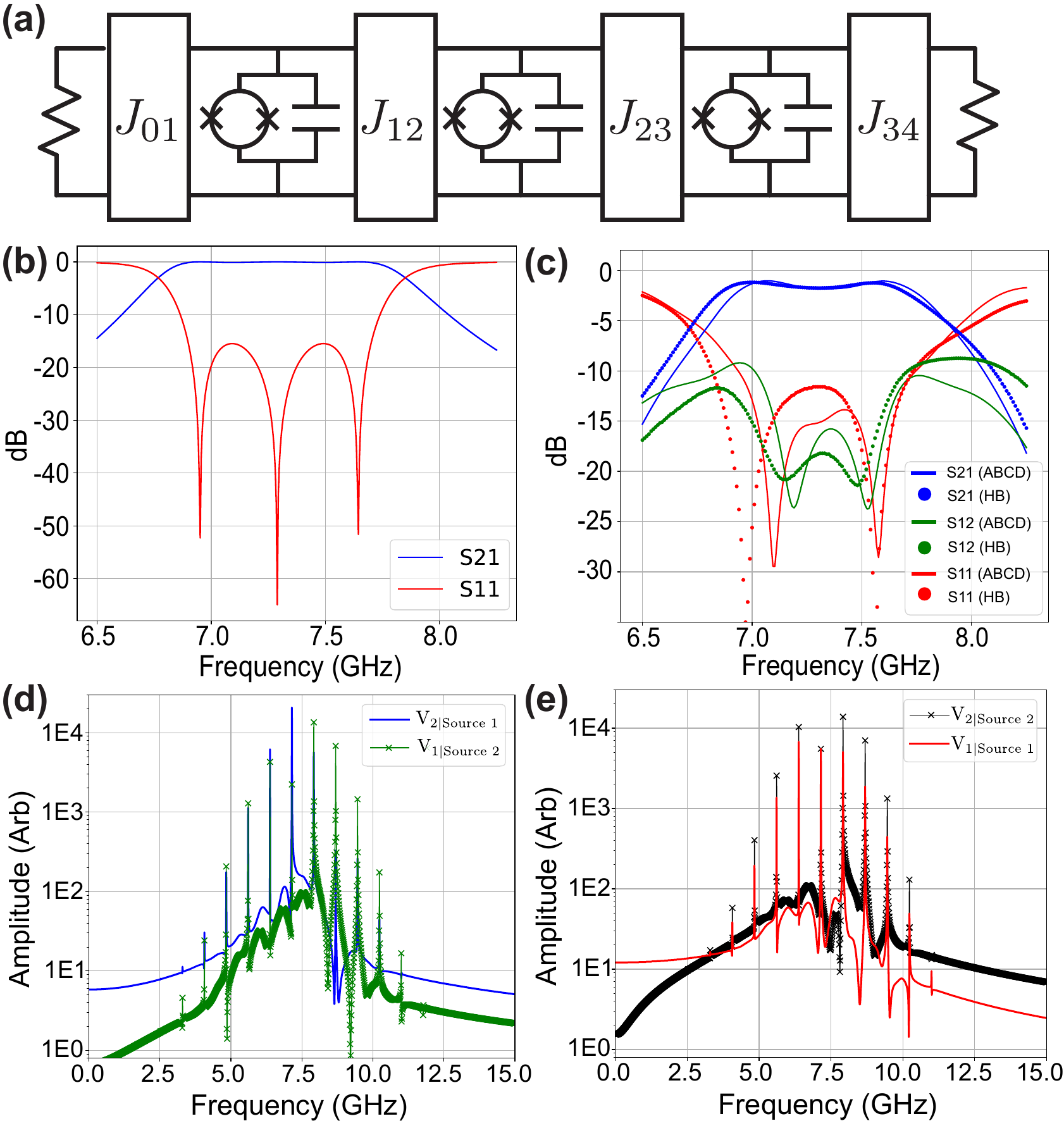}
      \caption{(color online) \textbf{(a)} Three pole admittance inverting filter with DC--SQUIDs as shunt inductors. The admittance inverters $J_{XY}$ connect the modes of the filter and can be implemented as capacitor or inductor $\pi$--networks or as 90$^\circ$ transmission lines. \textbf{(b)} ABCD matrix calculated filter response with DC flux applied to each SQUID with no RF modulation. \textbf{(c)} Comparison between the ABCD matrix calculation and harmonic balance simulation of the filter response with both DC bias and RF modulation of the DC--SQUIDs. For the harmonic balance simulation, ideal inverters were replaced with capacitive $\pi$--networks. With degenerate frequency RF pumps applied to each of the three SQUIDs, tuning of the discrete phase differences between them can produce asymmetric transmission $|S^{00}_{21}(\omega)| \neq |S^{00}_{12}(\omega)|$. An important characteristic of this design is that the filter stays matched with an in--band return loss greater than 10 dB. \textbf{(d)} Calculated power spectra of the time domain simulated transmitted voltage for the circuit in \textbf{(a)} driven at 7.150 GHz. When driving port 1 and measuring at port 2 ($V_{2|\text{Source 1}}$), the energy remains in the central frequency peak while in the reverse operation ($V_{1|\text{Source 2}}$), the energy is dispersed into the three--wave mixing sidebands. \textbf{(e)} Calculated power spectra of the time domain simulated reflected voltage spectrum of the circuit in \textbf{(a)} when driven at 7.150 GHz.}
      \label{Fig:ABCD_Scat_Params}
   \end{figure}

A proposed physical multi--pole filter topology is illustrated in FIG.~\ref{Fig:ABCD_Scat_Params}(a). It is a three--pole admittance inverting filter where the shunt inductor of each pole is replaced with a DC-SQUID. The calculation of the implemented isolating filter response is performed via cascading of the ABCD matrix for each admittance inverter and subsequent shunt LC pole. The forms of the spectral ABCD matrices are calculated following \cite{Wu:2019aa} and shown here for a DC-SQUID. %
For a three--pole device, the ABCD matrix takes the form
\begin{equation}
    \resizebox{0.42\textwidth}{!}{$
    \overline{\mathbf{A}}_\text{3P} = 
        \renewcommand\arraystretch{2}
        \begin{bmatrix}
            \frac{J_{34}J_{12}^2 \mathbf{I}_{n} + J_{34}\overline{\mathbf{Y}}_\text{P1} \overline{\mathbf{Y}}_\text{P2}}{J_{01}J_{12}J_{23}} & \frac{J_{23}^2\overline{\mathbf{Y}}_\text{P1} + J_{23}^2 \overline{\mathbf{Y}}_\text{P3} + \prod\limits_{a=1}^{3}\overline{\mathbf{Y}}_\text{Pa}}{J_{12}J_{23}J_{34}} \\
            \frac{J_{01}J_{34}\overline{\mathbf{Y}}_\text{P2}}{J_{12}J_{23}} & \frac{J_{01}J_{23}^2 \mathbf{I}_{n} + J_{01} \overline{\mathbf{Y}}_\text{P2}\overline{\mathbf{Y}}_\text{P3}}{J_{12}J_{23}J_{34}}
        \end{bmatrix} \, .$}
\end{equation}
%
We numerically calculate the response of a three--pole admittance inverting filter under both quiescent DC flux bias and RF modulation conditions. We choose a Chebyshev filter response with center frequency ${F_c = 7.3} \, \text{GHz}$, ${\text{BW} = 800 \text{MHz}}$, and in--band ripple ${r_\text{dB} = 0.125}$ dB. The impedance of the three poles were set to 15, 10, and 15 Ohms, respectively. For these simulations, we focus on the $\pm\omega_m$ and $\pm2\omega_m$ modulation products. With the addition to the signal tone, this results in 5$\times$5 spectral S--parameter matrices. Figure \ref{Fig:ABCD_Scat_Params}(a) displays a schematic of the simulated circuit. For no RF modulation and when biased at a value of $\beta = 0.3\pi$, the ABCD matrix calculated response of the filter is that of a standard band--pass filter as displayed in FIG. \ref{Fig:ABCD_Scat_Params}(b). 

We calculate the circuit response again but now with each SQUID pumped at a common RF frequency and RF amplitude, but with differential phases. To verify the methodology, we compare the ABCD matrix calculated results with that of a numerical harmonic balance \cite{Steer:1991} simulation of the circuit. For the harmonic balance simulations, equivalent capacitive $\pi$--networks were substituted to implement the admittance inverters and a mathematically equivalent non--linear model of the DC--SQUID was utilized. In effort to capture the full non--linearity of the DC-SQUID, 9 mixing orders in the signal and pump frequencies were utilized. Figure \ref{Fig:ABCD_Scat_Params}(c) displays the overlaid results of both the spectral ABCD matrix calculation and the harmonic balance simulation. For the spectral ABCD matrix calculation, $\beta = 0.3\pi, \, \alpha = 0.064\pi$ and a pump frequency $f_m = 691$ MHz were used. The phases of the pumps were set to 0, $\pi/4$, and $\pi/2$, respectively. To achieve a matching response in the harmonic balance simulations, the pump frequency and RF pump amplitude had to be modified slightly to $f_m = 770$ MHz and $\alpha = 0.044\pi$, respectively. The modification of these parameters to achieve a phenomenologically matching response is expected as in the ABCD calculation, the DC--SQUID non--linearity is only expanded to 2nd order. Higher orders of the applied flux $\mathcal{O}(\Phi_A^n)$ where $n>2$ provide small but non--negligible alterations to the filter response that are more accurately captured in the harmonic balance simulations.

We note again that the filter networks are employed to integrate the SQUIDs into a  $Z = 50\, \Omega$ environment at DC flux bias \textit{only} ($\alpha = 0$). From the previous treatment of the DC--SQUID non--linearity, when RF flux bias is applied ($\alpha \neq 0$), corrections to the bare SQUID inductance arise. However this approximation of treating the circuit linearly to first couple and match the SQUIDs and treating the non-linear modulation as a perturbation to that match allows us to separate the matching problem of the SQUID(s) to both ports and the non-linear modulation of the circuit.  By choosing a topology that allows the overall circuit with the SQUID to remain matched we can focus on optimizing the non-linear modulating phase for non-reciprocal transmission as shown in Section~\ref{section:Circuit}. As shown in FIG.~\ref{Fig:ABCD_Scat_Params}(c), this is a good design approximation.  However, it is important to note that the RF flux bias does slightly degrade the filter match, effectively reducing the number of poles in the matching network (see FIG. \ref{Fig:ABCD_Scat_Params}(b--c)). We reserve the employment of more sophisticated matching networks to maintain the match while under RF modulation for future work and note that the incorporation of an additional matching network could have further improvement on the bandwidth and/or non-reciprocal behaviour \cite{Naaman:2021}.

Figures \ref{Fig:ABCD_Scat_Params}(d) and \ref{Fig:ABCD_Scat_Params}(e) display the calculated power spectra derived from time--domain simulation results for the reflected and transmitted voltage spectra from the three--pole isolating filter \textcolor{black}{at the signal frequency at 7.150 GHz, where the directionality was at a maximum}. In the forward direction (drive port 1 and measure at port 2), the voltage is three--wave mixed from the carrier frequency and back resulting in constructive interference and near unity transmission. The additional insertion loss is due to the imperfect mode conversion back to the original signal frequency resulting in residual voltage in the sidebands. This residual voltage is then either reflected back out of port 1 (FIG. \ref{Fig:ABCD_Scat_Params}(e), red curve) or transmitted through to port 2. In the reverse direction (drive port 2 and measure at port 1), the power at the signal mode is three--wave mixed \textit{out of band} of the filter and reflected back out port 2 (Fig \ref{Fig:ABCD_Scat_Params}(e) Black Curve). Due to the combination of the filter's finite roll--off and imperfect mode conversion, the achieved isolation is not infinite, with residual voltage at the signal frequency remaining unconverted and thus transmitted to port 1 with subsequent sidebands (FIG. \ref{Fig:ABCD_Scat_Params}(d), green curve).  

Overall, the asymmetric transmission ratio at the signal frequency is maximized in the linear filter band. In general, the harmonic balance simulation and spectral S--parameter model show good agreement for both the transmitted and reflected S--parameters at the signal frequency. While the harmonic balance model captures a more detailed picture of the non--linear behaviour of the DC--SQUID, the good agreement shows that the physics of the circuit are well captured via the spectral S--parameter calculations demonstrating that the lowest order modes dominate the response of the circuit when under modulation.  Another important aspect of the device simulations is that they show the device remains matched under RF modulation with an in--band return loss greater than 10 dB.

\section{Design and Fabrication}\label{Section:Fab}
\begin{figure*}[t]
    \centering
      \includegraphics[width = \textwidth]{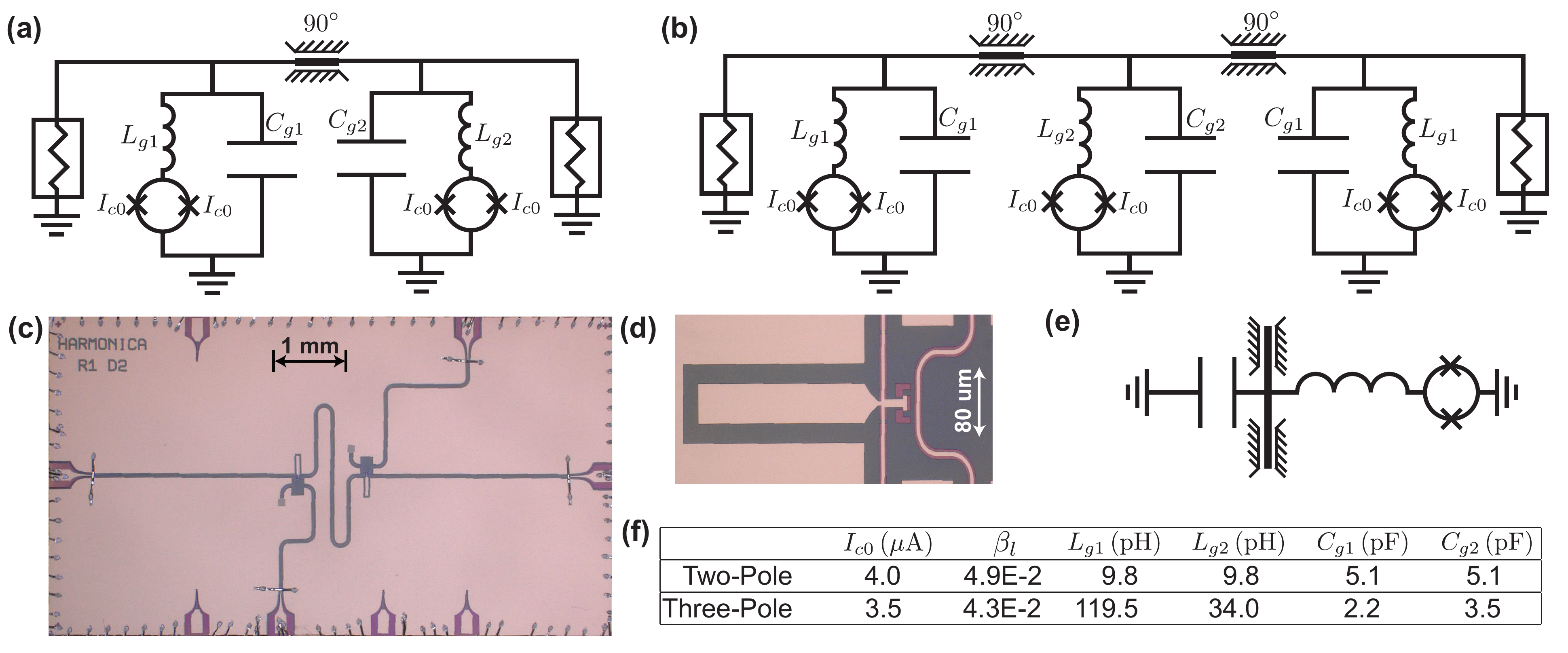}
      \caption{(color online) \textbf{(a)} Schematic of the fabricated two--pole device. \textbf{(b)} Schematic of the fabricated three--pole device \textbf{(c)} Micrograph of the two--pole device \textbf{(d)} Micrograph of a LC pole shunt capacitor and DC--SQUID. \textbf{(e)} Circuit schematic detailing the layout shown in \textbf{(d)}. \textbf{(f)} Values for the circuit elements shown for the two and three--pole devices in \textbf{(a)} and \textbf{(b)}, respectively.}
      \label{Fig:wiringDiagram}
\end{figure*}
To experimentally verify the above theory, we fabricated two different circuit designs utilizing two-- and three--pole filter structures respectively. In each design, the impedance of the poles was determined so as to allow for $Z = 50 \Omega$ $90^\circ$ transmission line couplers between the poles (See Appendix \ref{TxApp} for more details).
A common DC--SQUID geometry was employed in all designs with a geometric inductance of 12.8 pH as extracted with InductEx \cite{Fourie:2011aa}. For each pole, a portion of the geometric shunt inductance is replaced by its flux--biased DC--SQUID equivalent, the proportion of which was numerically optimized to provide nominally identical critical currents in excess of $I_c > 1 \, \mu$A. The component parameters for the two devices are listed in FIG.~\ref{Fig:wiringDiagram}(f).
For the two--pole design, the poles are degenerate in their respective component values while for the three--pole design, the first and third pole each employ the same component values thus allowing for the geometric inductance and capacitance of poles one ($L_{g1} / C_{g1}$) and three ($L_{g3} / C_{g3}$) to be equal. Figures \ref{Fig:wiringDiagram}(a) and \ref{Fig:wiringDiagram}(b) display the schematics for the two-- and three--pole circuit with the values for those respective circuit elements listed in FIG. \ref{Fig:wiringDiagram}(f).

The devices are fabricated utilizing a Nb tri--layer process with a critical current density $J_c=6\, \mu \text{A}/\mu \text{m}^2$ \cite{Cui:2017} . The process is comprised of two superconducting metallic layers with an insulating TEOS SiO$_x$ dielectric layer. The Nb base and counter electrode thicknesses were each 200 nm with the interstitial SiO$_x$ planarized to a thickness of 100 nm between the two metal layers. All layers are patterned atop a thermally oxidized Silicon substrate. The base Nb layer serves as a ground plane for the circuit while the second niobium layer allows for contact to the JJ counter electrode and additional on--chip wiring. 
A micrograph of the two--pole circuit is displayed in FIG.~\ref{Fig:wiringDiagram}(c).  Each DC--SQUID has its own independent flux bias line (bottom left / top right) allowing for the application of DC and RF flux signals. A zoomed micrograph of an individual filter pole is presented in FIG.~\ref{Fig:wiringDiagram}(d). The corresponding schematic detailing the parallel LC pole layout is shown in FIG.~\ref{Fig:wiringDiagram}(e)

\section{Measurements}\label{section:Measurements}

Packaged devices were loaded in a dilution refrigerator and thermally sunk to the mixing chamber stage.
Cryogenic through--reflect--line (TRL) \cite{Ranzani:2013} standards were implemented for coaxially de--embedding the cabling, attenuation, and amplification in--between the VNA and the DUT up to the devices SMA cables. Because of the type of connector (Ardent TR multicoax) used to connect the device package, approximately 18" of residual coaxial cable (flexible 047 coaxial cable) at either port was unable to be de--embedded, along with the package PCB and wirebonds.  The additional insertion loss from these elements are thus included in the measured insertion and return loss.  

An explicit wiring diagram FIG.~\ref{Fig:FridgeDiagram} is included in Appendix \ref{APP:Wiring}. The RF input lines to the TRL setup were attenuated at the 4K and mixing chamber stages for a total of 66 dB of explicit attenuation at the input. The output amplification is provided via cryogenic low--noise amplifiers at the 4K stage cascaded with low--noise amplifiers at room--temperature.  The flux biases are delivered via coaxial lines with 20 dB of attenuation at the 4K stage and low--pass filtering at the MXC stage to protect the device from out--of--band noise. The room--temperature based signals are generated via DC current sources (Yokogawa GS200) and an RF continuous--wave source (Holzworth HS900A) combined via a bias--tee at room--temperature. The device was characterized by a vector network analyzer (Keysight N5242B).

\subsection{Two--Pole Isolating Filter Response}

\begin{figure}
    \centering
      \includegraphics[scale=0.4]{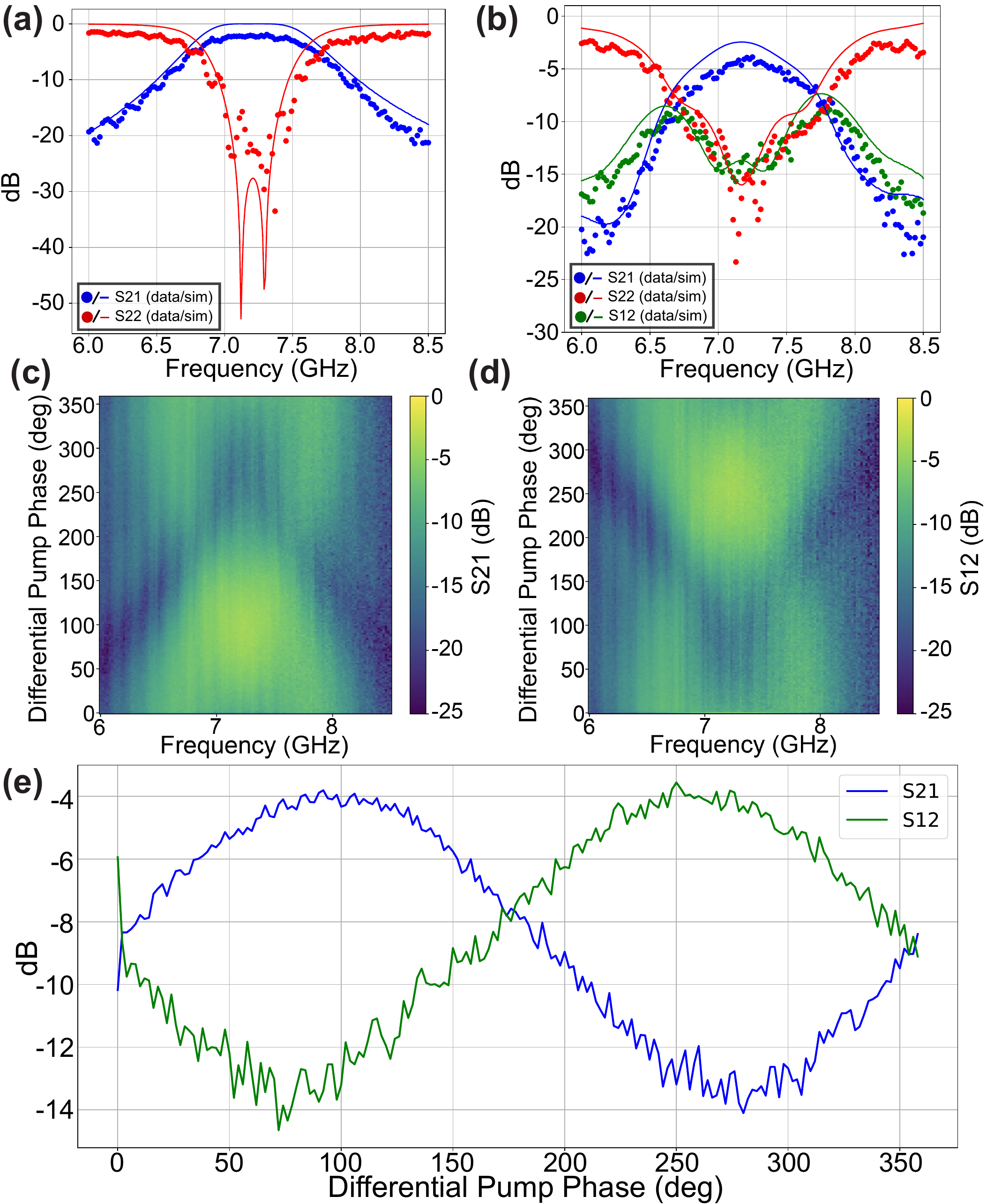}
      \caption{(color online) \textbf{(a)} Simulated and measured response of the two--pole circuit at DC Flux bias. \textbf{(b)} Simulated and measured response of the two--pole circuit under RF modulation of the DC--SQUIDs. \textbf{(c--d)} $S^{00}_{21}$ and $S^{00}_{12}$ as a function of the differential pump phase and applied signal frequency. \textbf{(e)} Vertical line cuts of \textbf{(c)} and  \textbf{(d)} at $f = 7.2$ GHz. The data shows that by adjusting the differential pump phase between the two DC--SQUIDS, the asymmetric transmission can be reversed from port 1 to 2 and vice versa.}
      \label{Fig:R1B2_Data}
\end{figure}
The measured DC flux biased response of the filter with accompanying harmonic balance simulation results are displayed in FIG. \ref{Fig:R1B2_Data}(a). An additional 2 dB of insertion loss is seen in the data when compared to the simulation. We attribute this difference to the imperfect calibration of the sample whereby the reference planes of the two--port calibration were offset from that of the sample planes by the additional coaxial cable, PCB traces, and wirebonds at either RF port (see Appendix \ref{APP:Wiring} for an experimental wiring diagram).
Despite this limitation, there is still excellent agreement between the simulation and measured data both in terms of bandwidth of the filter and the in--band return loss level.

The measured flux pumped response of the two--pole isolating filter is shown in FIG. \ref{Fig:R1B2_Data}(b) along with corresponding harmonic balance simulation of the device. The applied pump frequency was set to $f_m = 670$ MHz. Accounting for only the explicit attenuation in the flux line, the experimental RF flux pump amplitude $\alpha_\text{E} = 0.065\pi$. The shown simulation results utilized a RF flux pump amplitude $\alpha_\text{S} = 0.040\pi$. Again we find excellent qualitative agreement between measurement and simulation. The difference in the max forward transmission between the measured and simulated response is $\sim$ 3 dB. The majority of this discrepancy can again be explained by the additional insertion loss from the physically offset calibration planes. The additional $\sim$ 1 dB of insertion loss seen while RF pumping we attribute to higher order inductance offset terms arising from the RF pump amplitude which effects the impedance mismatches of the device.
\begin{figure}[]
    \centering
      \includegraphics[width = 0.43\textwidth]{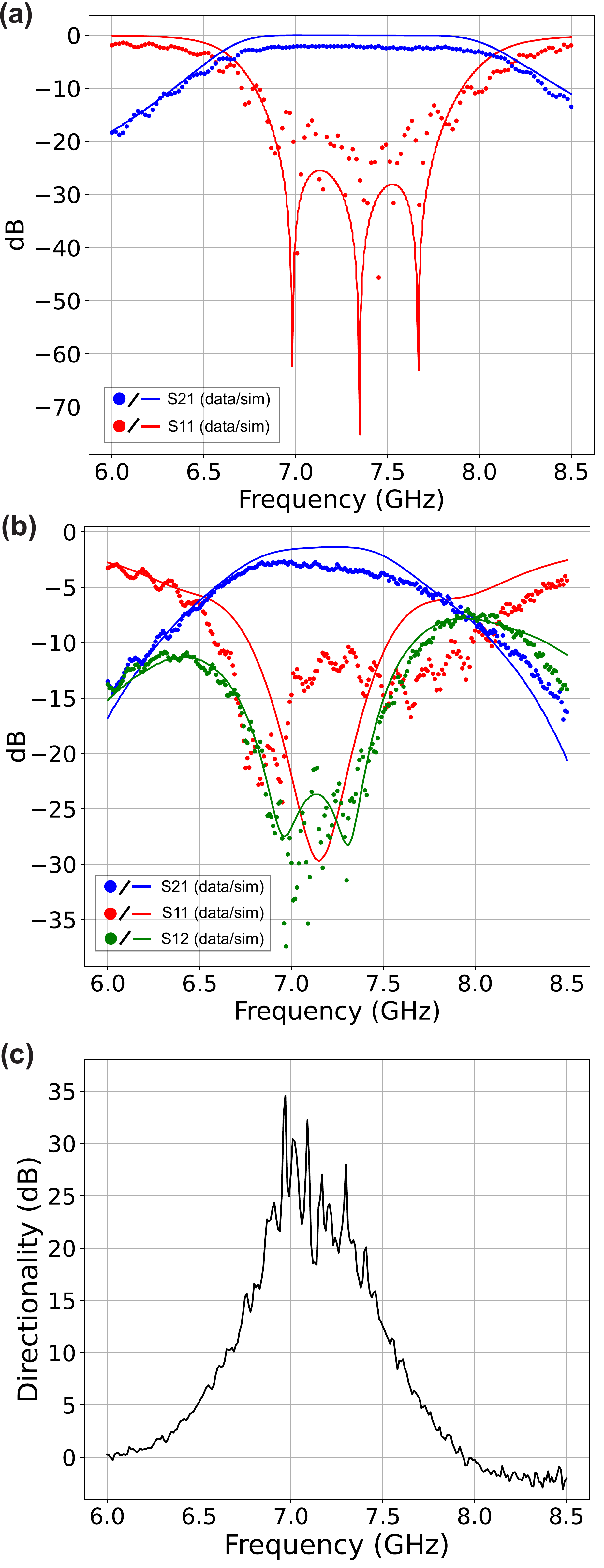}
      \caption{(color online) \textbf{(a)} Measured and simulated response for a three--pole filter at optimized DC flux bias. \textbf{(b)} Measured and simulated three--pole filter response when pumped at $f_p = 1100$ MHz with differing phases between pumps. The simulated response had differential phases between the three pumps of 0, 45.15, and 88.57 degrees, respectively. \textbf{(c)} Measured directionality $\text{D} = |S^{00}_{21}|/|S^{00}_{12}|$ of the three--pole isolating filter. }
      \label{Fig:R1B3_Data}
\end{figure}
The phase offset between the two RF flux pumps in simulation was set to $\Delta \phi = 77.4^\circ$. Experimentally the absolute phase difference between the pumps at chip--level could not be determined due to differing electrical lengths between the independent pump channels. The appropriate phase offset between the signals was determined experimentally.
\begin{figure}[b]
    \centering
      \includegraphics[width = 0.46\textwidth]{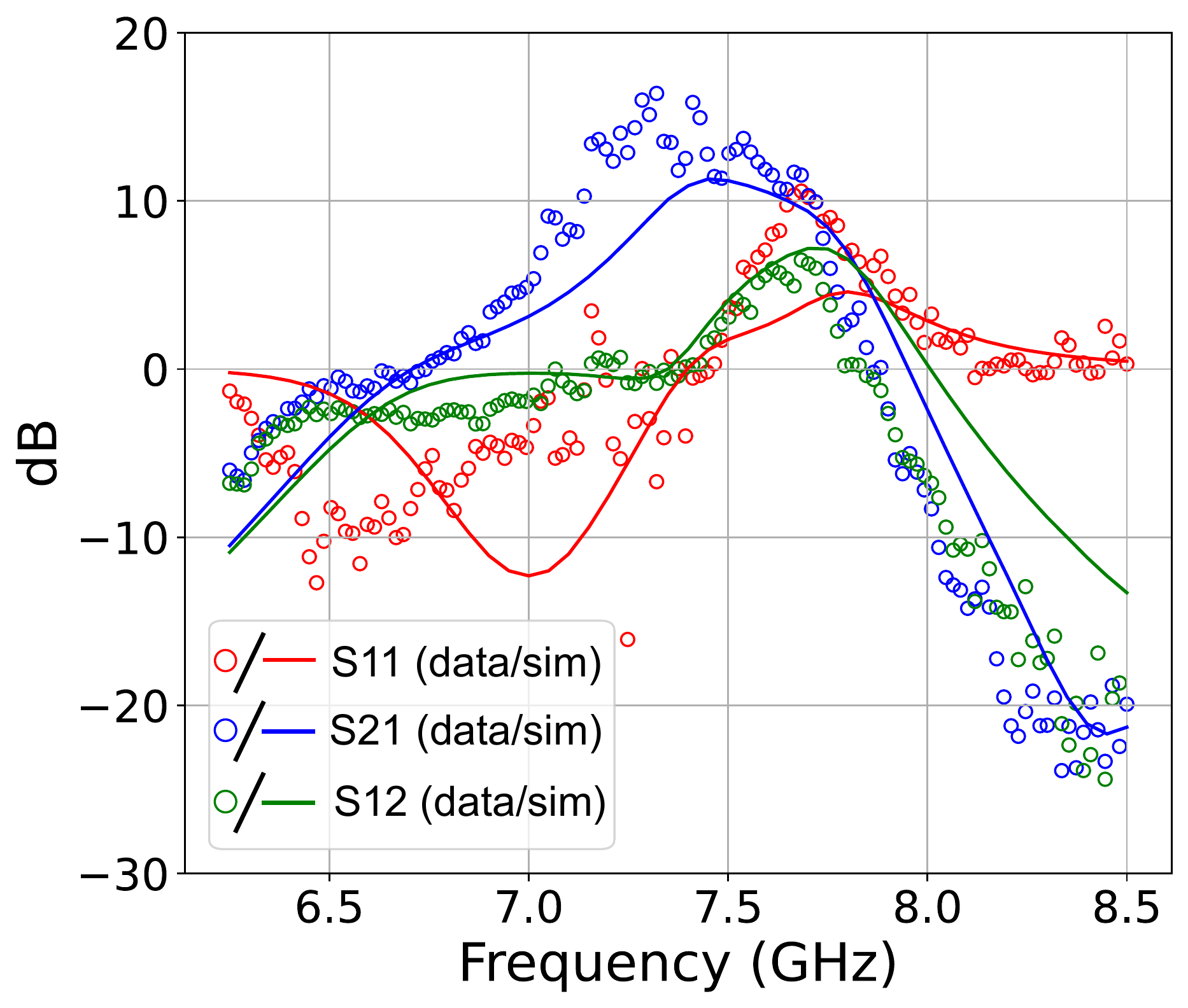}
      \caption{(color online) Measured and simulated scattering parameters for the 3 pole device when pumped at 14.69 GHz.}
      \label{Fig:3PoleAmpFig}
\end{figure}
\begin{figure*}[t]
    \centering
      \includegraphics[width = \textwidth]{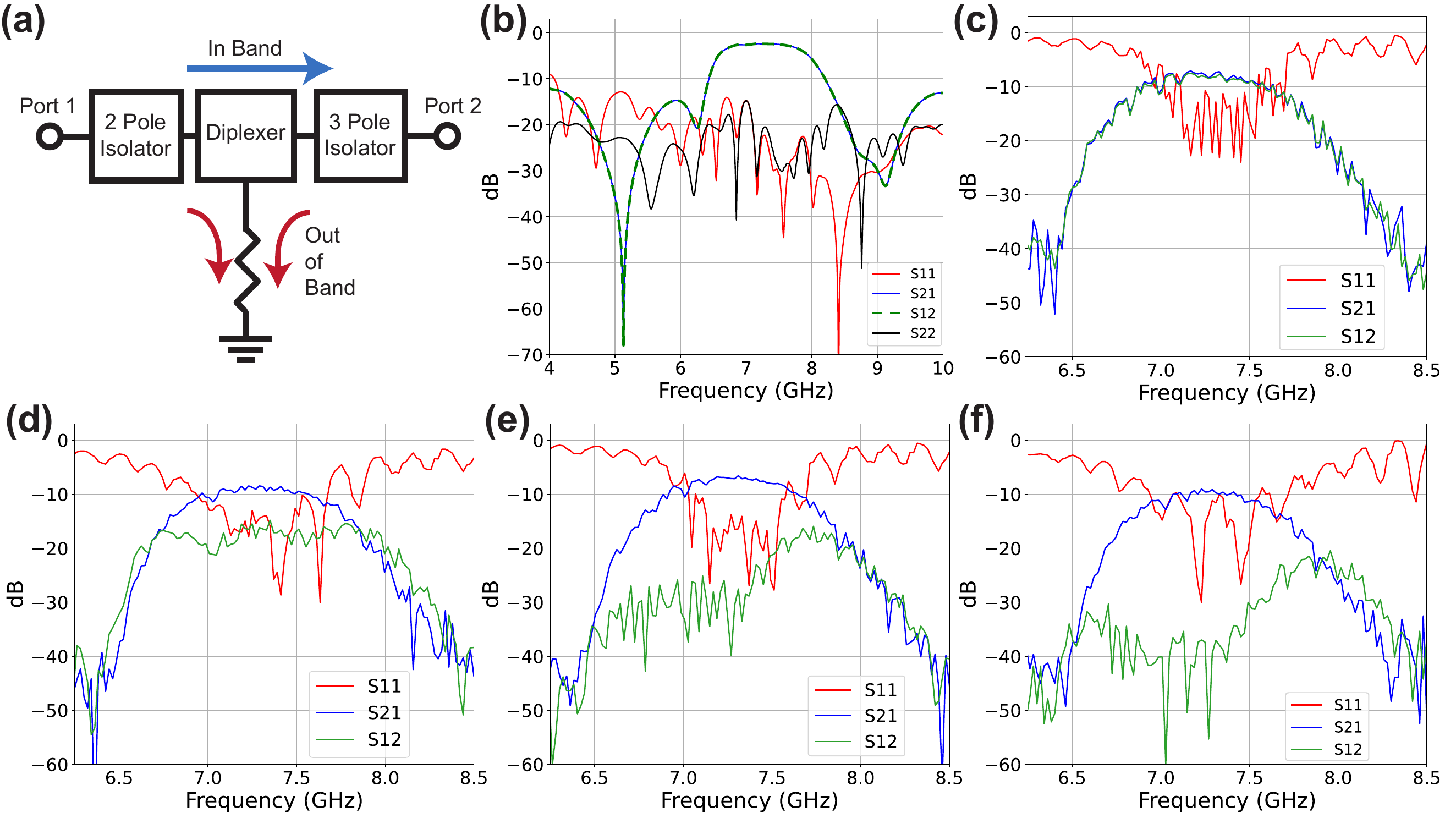}
      \caption{(color online) \textbf{(a)} Circuit block diagram detailing how the isolator devices were cascaded. 
      \textbf{(b)} Measured room temperature scattering parameters of the intermediate diplexer element at the ports used to connect the two isolator blocks. \textbf{(c)} DC flux tuned up cascaded isolators. The insertion loss of 8 dB is a combination of the insertion loss from the two filters and the diplexer plus an additional of 72 inches of microwave cable that could not be calibrated out of the measurement. \textbf{(d)} Cascaded isolator response with RF pumps applied solely to the two--pole device. 
      \textbf{(e)} Cascaded isolator response with RF pumps applied solely to the three--pole device. 
      \textbf{(f)} Cascaded isolator response with RF pumps applied to both the two and three--pole devices.}
      \label{Fig:CascadedIsolators}
\end{figure*}
Figures~\ref{Fig:R1B2_Data}(c) and \ref{Fig:R1B2_Data}(d) display the measured forward and reverse scattering parameters of the two--pole device as a function of the differential phase offset. Figure \ref{Fig:R1B2_Data}(e) displays resulting vertical line cuts through \ref{Fig:R1B2_Data}(c) and \ref{Fig:R1B2_Data}(d) at f = 7.2 GHz. This data clearly shows that the direction of power flow in the device can be set by adjusting the differential phase between of the pumps as predicted in FIG.~\ref{Fig:MInverse}(c).

\subsection{Three--Pole Isolating Filter Response}

The measured and simulated responses for the three--pole device under quiescent DC flux bias and DC plus RF modulation are displayed in FIG.~\ref{Fig:R1B3_Data}(a) and FIG.\ref{Fig:R1B3_Data}(b), respectively. Referring to FIG. \ref{Fig:R1B3_Data}(a), we see excellent agreement between the harmonic balance simulation and the measured data at quiescent DC flux bias. There is $\sim$ 2 dB discrepancy between the in--band through levels reported by the simulation and the measured data which we ascribe again to the imperfect calibration mentioned earlier. The RF pumped isolating filter response is displayed in FIG.~ \ref{Fig:R1B3_Data}(b) where the modulation tone was set to ${f_m = 1100\, \text{MHz}}$. The estimated on--chip RF pump amplitude ${\alpha_\text{E} = 0.078\pi}$. The corresponding simulation was performed with a pump frequency $f_p = 1100$ MHz, pump amplitude ${\alpha_\text{S} =  0.09\pi}$ and differential phases between the three pumps of 0, 45.15, and 88.57 degrees, respectively. The phase offsets required to produce the measured spectrum were found experimentally. When properly set, we experimentally realize an isolating band--pass filter with and insertion loss ${\text{IL} < 5 \text{dB}}$ and a directionality ${\text{D} > 15\, \text{dB}}$ over a bandwidth ${\text{BW} = 600}$ MHz. We note that the measured $S_{11}^{00}$ degrades mid--band as compared to the simulation. However the measured circuit does stay matched with a response better than -10 dB over the entire bandwidth. The directionality D of the device is plotted in FIG.~\ref{Fig:R1B3_Data}(c) From 6.9--7.5 GHz, we achieve a directionality in excess of 15 dB.

\subsection{Directional Amplification}
An independent mode of operation for the circuits described above is that of directional amplification. This mode is achieved by placing the pump frequency far above the center frequency of the pass band $f_m \gg f_c$. Setting $f_m = 2f_c$, the down converted terms in the upper--half of the impedance matrix in Eq.~\eqref{eq:squidZ} develop a negative and real amplitude (for the proper pump phase) yielding the potential for gain. The key to achieving gain directionality is again the provision of two or more pumps with different phases such that the amplified idler modes can be mixed back to the signal mode constructively in one direction and destructively in the other.

To probe this mode of operation, we apply modulation tones of frequency ${f_m = 14.69 \text{ GHz}}$ to each SQUID of the three--pole device. The on--chip amplitude for each pump tone is calculated to be ${\alpha_E = 0.099\pi}$. The resulting measured and simulated frequency response for $S^{00}_{11}$ (red), $S^{00}_{21}$ (blue), and $S^{00}_{12}$ (green) are displayed in FIG. \ref{Fig:3PoleAmpFig}. Both theory and measurement demonstrate that the device remains passive in reflection ($S^{00}_{11}$ $<$ 0) up to 7.3 GHz while simultaneously showing gain in the forward direction, which is a different mode of operation than traditional reflection based amplifiers. With regards to the forward scattering parameter $S^{00}_{21}$, we measure greater than 5 dB of gain over a bandwidth of ${\text{BW} \approx 700 \text{ MHz}}$ with a peak gain of 15 dB at 7.25 GHz. Conversely, ${S^{00}_{12} < 0\,  \text{dB}}$ up to approximately 7.4 GHz after which we begin to measure reverse amplification between 7.4--7.9 GHz with a maximum of 6 dB of gain at 7.7 GHz. In all measured parameters, we see very good model--hardware correlation. 

We wish to emphasize here that the measurement of amplifying capabilities of the circuit were done as a verification of the models. To achieve broadband directional amplification, further research, design, and development is required. With that, expansion into the calibration of standard amplifier specifications including added noise and saturation power is beyond the scope of this work.

\subsection{Cascaded Filters}
With the demonstration of the two-- and three--pole isolating filters above, an obvious future path towards increasing the directionality for a set bandwidth is to increase the number of poles. While we leave that direction for future exploration, it is worth noting some of the difficulties with that approach. Specifically, moving to an $n>3$ pole filter topology requires a more complicated device bring up and operation. Specifically, beyond just considering fabrication variation, this includes DC bias point optimization and static phase offsets at each DC--SQUID that must be optimized and maintained across the device.

A notable way to alleviate this complexity is to cascade two or more multi--pole devices in effort to achieve larger directionality. We highlight this approach for a few reasons. First, in comparison to building a more complex circuit, the operation of two cascaded circuits can result in a simpler bring up. Each cascaded device can be brought up independently with respect to DC bias and pump phase offsets. With little to no interplay between the respective biases and pumps between devices, the requirements on the phases of the pumps are simpler to implement. Second, cascading parametric devices is a critical path to achieving tightly integrated, small footprint, readout electronics required for future quantum processors. Finally, in absence of large environmental offsets or fabrication variation, identical devices, once cascaded, ideally can share the same bias and pump lines significantly reducing wiring overhead.

To demonstrate this idea, we present data obtained from experiments on the previously characterized two-- and three--pole devices cascaded in series with an intermediate microwave diplexer. The basic experimental setup for these experiments is shown in FIG.~\ref{Fig:CascadedIsolators}(a). The diplexer acts as an impedance match for the larger circuit across the full band of interest (including the stopband under linear operation). The diplexer performs the function of filtering out and directing the sidebands that arise from both the two--pole or three--pole device while under modulation into a 50 ohm termination while simultaneously allowing the signals in the pass band to travel from port 1 to port 2. By filtering out generated sidebands between devices we prevent any unwanted standing waves and allow for each device to be treated as an independent circuit. The diplexer is constructed using discrete commercial off--the--shelf components (Krytar Hybrid Coupler 3060200, Mini--circuit filters, VLF--6000+ and VHF--8400+ and cryogenic 50--Ohm terminations from XMA Corporation). The diplexer's S--parameters, as measured at room temperature, are displayed in FIG.~\ref{Fig:CascadedIsolators}(b). We note that this passive circuit could be incorporated on--chip or on--package with small overhead in the overall design.

FIG.~\ref{Fig:CascadedIsolators}(c) displays the S--parameters for when both filters are biased with DC flux. The bandwidth of the pass--band is set by the two--pole filter. The best obtained insertion loss for the entire circuit was 8 dB. This insertion loss with respect to the calibration plane results from a combination of the insertion loss from the filters, diplexer, and an additional 72 inches of microwave cable required to wire the experiment due to the connectors on both devices (Ardent TR multicoax). Both circuits were then tuned up independently for asymmetric transmission via application of RF flux to their respective LC poles. FIG.~\ref{Fig:CascadedIsolators}(d) displays the isolating filter response for when only the two--pole filter SQUIDs are pumped. The resulting directionality is ${\text{D} \sim 8}$ dB between 7.0--7.5 GHz. FIG.~\ref{Fig:CascadedIsolators}(e) displays the cascaded filter response with the RF pumps applied the three--pole device SQUIDs. A directionality $D \sim 20$ dB is obtained with negligible additional insertion loss as compared to the non--pumped response. Finally, the data when both isolators are pumped is presented in FIG. ~\ref{Fig:CascadedIsolators}(f). The additive effect of each filter's isolation is clearly seen in the suppression of the S12 over the band from 7.0--7.5 GHz, ultimately reaching a directionality in excess of 25 dB. Crucially, in all modes of operation, the device remains matched in the band of interest with an in--band return loss greater than 10 dB. The ability to cascade these parametric devices, despite their non--linear mode of operation, allows for a straight--forward method to incorporate these devices into a larger QPU readout chain.

\section{FUTURE OUTLOOK}\label{Section:Future}

The designs and corresponding data presented above represent an initial step towards towards the realization of a superconducting integrated circuit replacement for traditional ferrite--based isolators. As noted above, this is an active area of research and during the preparation of this manuscript, we became aware of the work in \cite{Kwende:2023} which reports on the use of a circuit using a modulated RF-SQUID which can act as a circulator to achieve directionality in a 200 MHz band

To continue this research, future designs will have to address a few key issues seen with these initial devices. These issues are expanded upon below.

\subsection{Increased Directionality}\label{directionality}

The best directionality achieved in this work at a single frequency is $\sim$30 dB. To replace modern commercial ferrite--based isolators, this level of directionality must be achieved across the entirety of the filter band. In order to accomplish this, two natural approaches are considered. The first and most straight--forward approach is the simple cascading of multiple isolator devices in series. As shown above, for this to properly work, interstitial filters must be placed between the isolator stages to filter harmonics generated by the circuit. While easy to implement, this approach has drawbacks including, but not limited to, the requirement of additional lumped element devices. These can, in theory, be integrated on--chip without too much overhead. The second approach would be the inclusion of more poles in the filter design so as to allow for more pumps and thus more mixing control. The main draw back in this approach is that the number of high bandwidth lines starts to become untenable. However, with the phases determined in advance from simulation, multiple pump lines could be combined in a single line with a deterministic phase delay implemented between sequential SQUIDS.

\subsection{Decreased Insertion Loss}\label{IL}

In terms of the readout requirements of QPUs, accumulating every measurement photon so as to maximize the quantum efficiency of the readout chain is extremely important \cite{Bultink:2018}. Any device inserted in the readout chain that adds loss will naturally reduce the quantum efficiency. The pumped devices presented above show additional insertion loss when compared to their non--modulated counterparts. For true integration, the insertion loss of these devices must be lowered. We reserve future work to explore more advanced methods in matching the SQUIDs not just at DC flux bias, but while under RF modulation.

\subsection{Control Wiring Overhead}

For these initial demonstrations, each SQUID in this work was driven via its own independently controlled flux bias line. This allowed for experimental differential flux offset compensation between all the SQUIDs and for the independent control of RF phases. For these devices to serve as viable alternative to current isolation devices, the flux bias control wiring overhead should be considered as a hurdle to overcome. A path towards achieving this is to engineer the circuit such that each DC--SQUID requires a common DC flux offset such that a common line can be utilized. Further, the circuit and common line can then be engineered together such that, for the chosen pump frequency, an appropriate electrical length between SQUIDs is allowed for.

\section{CONCLUSIONS}\label{Section:Conclusions}

In summary, we have designed, modeled, and tested a two--port non--reciprocal device utilizing the three--wave mixing capabilities of DC--SQUIDs embedded in multi--pole admittance inverting filters. Utilizing coupled--mode theory, we extracted limitations on the pump frequency, pump phase, and pump amplitude so as to achieve non--reciprocity that describe well work published in the literature. We also provide a straightforward model that shows how three-- and four--wave mixing arises from the inherent non--linear inductance of a DC--SQUID and can be utilized to realize non-reciprocal microwave transmission. This model was further verified via harmonic balance simulations where the full non--linearity of the SQUID inductance could be more properly captured. These models show excellent agreement with measured data taken on two-- and three--pole isolator devices where we have achieved over a bandwidth of 500 MHz an in--band directionality of 8 and 15 dB, respectively. Further, we show with the three--pole device that directional amplification in excess of 10 dB over a 500 MHz BW can be achieved. Finally, we have presented measurements on the cascaded performance of a two-- and three--pole isolating filter where a directionality approaching 30 dB was obtained with minimal added insertion loss from the RF pumps.

\appendix

\section{TRANSMISSION LINE ADMITTANCE INVERTERS}\label{TxApp}

The input admittance inverter $J_{01}$ between a system impedance $Z_0$ and that of the first shunt LC resonator $S_{r1}$ of a multi-pole filter has the form
\begin{equation}
    J_{01} = \sqrt{\frac{\overline{\omega}}{g_0 g_1 Z_0 Z_{r1}}}\, ,
\end{equation}
where $\overline{\omega} = (\omega_2 - \omega_1)/\sqrt{\omega_2 \omega_1}$, $\omega_1$ and $\omega_2$ are the lower and upper knee frequencies of the filter, and $g_{0}$ and $g_1$ are the first two terms in the low pass filter coefficients. By explicitly setting $J_{01} = 1/Z_0$ and solving for $Z_{r,1}$, one arrives at
\begin{equation}\label{Zr1}
    Z_{r,1} = \frac{\overline{\omega} Z_0}{g_0 g_1} \, .
\end{equation}
This is the impedance one must set the first LC resonator of the filter to in order to allow for a direct $Z_0$ input to the filter with no explicit inverter structure. For completeness, we also show how one designs the inter--resonator coupling to also be $Z_0$. The form of the inter--pole admittance inverter is between pole $n$ and $n+1$ is defined as
\begin{equation}
    J_{n, n+1} = \overline{\omega} \sqrt{\frac{1}{g_n g_{n+1} Z_{r,n} Z_{r,n+1}}} \, .
\end{equation}
Again we set the admittance value $J_{n,n+1} = 1/Z_0$ and solve for $Z_{r, n+1}$ yielding
\begin{equation}\label{Zrnp1}
    Z_{r, n+1} = \frac{Z_0^2 \overline{\omega}^2}{g_n g_{n+1} Z_{r, n}} \, .
\end{equation}
Taking $n=1$ and the result of Eq. \eqref{Zr1}, we can reduce Eq. \eqref{Zrnp1} to
\begin{subequations}
\begin{align}
    Z_{r, 2} &= \frac{Z_0^2 \overline{\omega}^2}{g_1 g_{2} Z_{r, 1}} \\
    & = \frac{Z_0^2 \overline{\omega}^2}{g_1 g_{2} \frac{\overline{\omega} Z_0}{g_0 g_1}} \\
    &= \frac{Z_0 \overline{\omega} g_0}{g_2}
\end{align}
\end{subequations}
The above outlined process can be continued to fully determine the impedance of every LC network in the filter once a filter type, center frequency, bandwidth, and ripple have been determined.

\begin{figure*}[t]
    \centering
      \includegraphics[width = \textwidth]{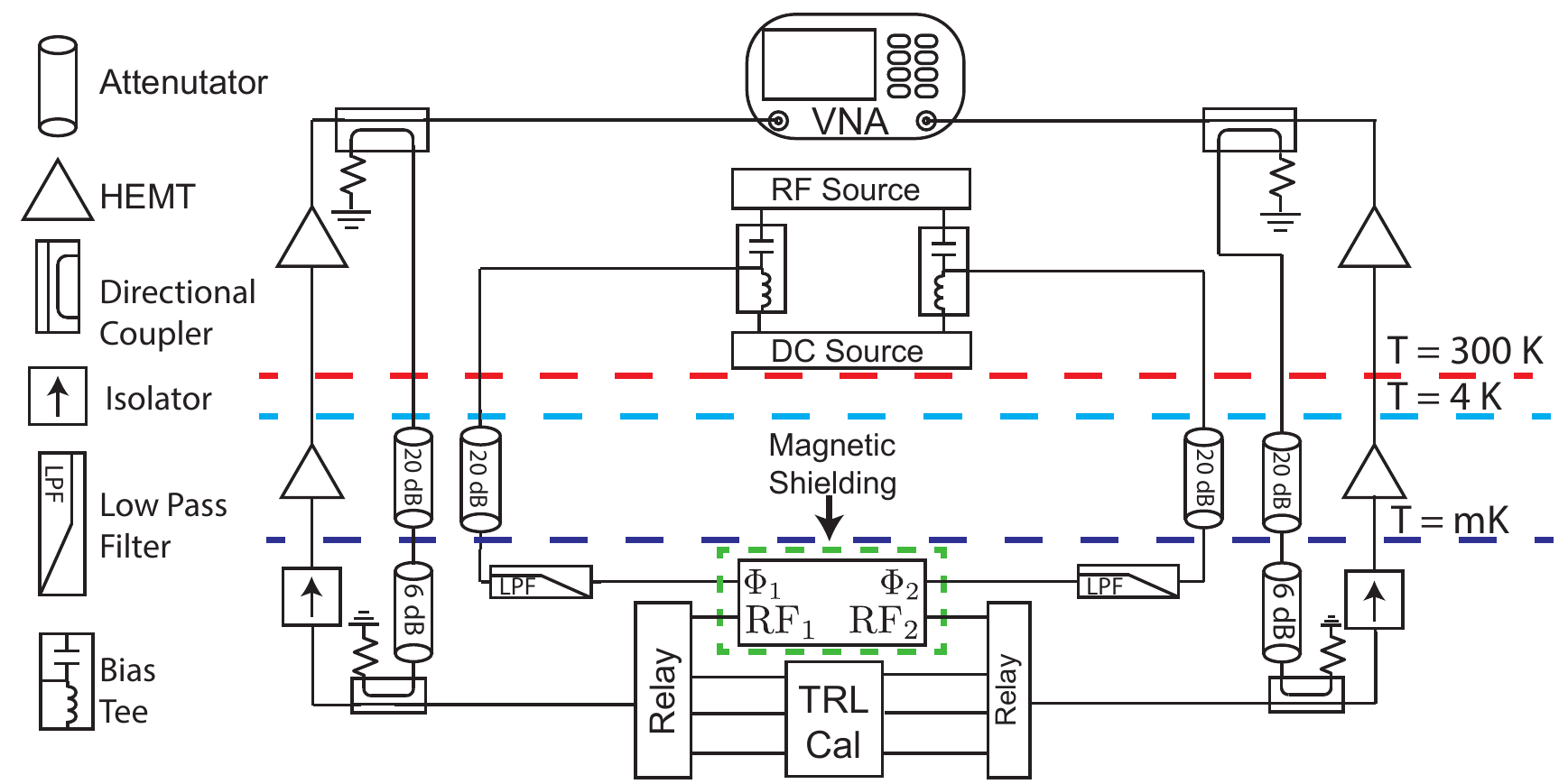}
      \caption{(color online) Experimental wiring diagram for the non--linear mixing isolator experiments. Here we should explicitly the wiring diagram for the two--pole device measurements. The experimental wiring was set such that the scattering parameter measurements could be calibrated both in transmission and reflection. One minor drawback to was the inability to calibrate directly at the input and output planes of the DUT. This inability with slightly increased insertion loss}
      \label{Fig:FridgeDiagram}
\end{figure*}

While the above treatment provides a method of determining the required impedance of each LC poles coupled via a transmission line of impedance $Z_0$, we note here that this treatment alone is not entirely sufficient as the electrical length of the transmission line must be explicitly defined. To calculate the appropriate electrical length of the transmission line, we reproduce here to the form of a transmission line of impedance $Z_0$ connected to a load of impedance $Z_L$ \cite{Pozar}

\begin{equation}\label{Ztx}
    Z_{\text{in}} = Z_0 \frac{Z_L + i Z_0 \tan (\kappa l)}{Z_0 + iZ_L \tan(\kappa l)} \, ,
\end{equation}
Where $\kappa = 2\pi/\lambda$ is the wave number, $\lambda$ is the wavelength, and $l$ is the electrical length of the transmission line. In the limit that $l \rightarrow \lambda/4$, the terms proportional to $\tan(\kappa l)$ begin to diverge leading to the simplification of Eq. \eqref{Ztx} with the form

\begin{equation}
    Z_\text{in} = \frac{Z_0^2}{Z_L} \, .
\end{equation}
Taking the simple transformations $Z_\text{in} = 1/Y_\text{in}$, $Z_0 = 1/Y_0$, and $Z_L = 1/Y_L$, we arrive at

\begin{equation}
    Y_\text{in} = \frac{Y_0^2}{Y_L} \, .
\end{equation}

\section{Experimental Wiring}\label{APP:Wiring}

\textcolor{black}{Figure}~\ref{Fig:FridgeDiagram} displays the in--fridge wiring utilized for the measurement of two--pole sample data. When measuring for the three-pole sample, an extra RF and DC source channel along with bias--tee was required. Channels 1 and 2 from the VNA were split via directional couplers such that transmission and reflections scattering parameters could be calibrated in--situ.

A minor drawback in this experimental setup was the inability to calibrate directly up to the input/output planes of the DUT. This experimental constraint manifested itself in measurement as additive insertion loss in both the forward and reverse scattering parameters stemming from the cabling between the relays and the RF channels of the DUTs.

\bibliographystyle{unsrt}
\bibliography{isolatorBib_2}
\end{document}